\documentclass{aastex631}

\usepackage{courier}
\usepackage{multirow}
\usepackage[inline]{enumitem}
\usepackage{amsmath}

\newcommand{\ticone}{TIC~219900027}
\newcommand{\tictwo}{TIC~229771234}
\newcommand{\tictwoM}{TIC~229771231}
\newcommand{\ticthree}{TIC~259006185}
\newcommand{\ticfour}{TIC~356896561}
\newcommand{\ticfive}{TIC~424461577}
\newcommand{\ticsix}{TIC~278956474}

\newcommand{\Solaris}{\textit{Solaris}}
\newcommand{\TESS}{\textit{TESS}}
\newcommand{\Gaia}{\textit{Gaia}}
\newcommand{\Kepler}{\textit{Kepler}}
\newcommand{\eleanor}{\texttt{eleanor}}
\newcommand{\Eleanor}{\texttt{Eleanor}}
\newcommand{\tesscut}{\texttt{TESScut}}
\newcommand{\astroquery}{\texttt{astroquery}}

\newcommand{\ie}{i.e.\ }
\newcommand{\eg}{e.g.\ }
\newcommand{\kvw}{KvW}
\newcommand{\mitnyan}{M24}
\newcommand{\mbf}[1]{#1}
\newcommand{\tbf}[1]{#1}
\newcommand{\boldm}[1]{#1}

\graphicspath{{./}{figures/}}

\begin{document}

\title{Precision timing of eclipsing binaries from TESS full frame images. Method and performance}

\correspondingauthor{Frédéric Marcadon}
\email{frederic.marcadon@villanova.edu}

\author[0000-0002-7606-7733]{Frédéric Marcadon}
\affiliation{Department of Astrophysics and Planetary Sciences, Villanova University \\
800 East Lancaster Avenue, Villanova, PA 19085, USA}

\author[0000-0002-1913-0281]{Andrej Pr\v{s}a}
\affiliation{Department of Astrophysics and Planetary Sciences, Villanova University \\
800 East Lancaster Avenue, Villanova, PA 19085, USA}



\begin{abstract}
Several hundreds of thousands of eclipsing binaries (EBs) are expected to be detected in the \textit{Transiting Exoplanet Survey Satellite} (\TESS{}) full frame images (FFIs). This represents a significant increase in the number of EBs available for eclipse timing variation studies. In this paper, we investigate the feasibility of performing precise eclipse timing of \TESS{} EBs using the FFIs. To this end, we developed a fast, automated method and applied it  to a sample of $\sim$100 EBs selected from the Villanova \TESS{} EB catalog. Our timing analysis resulted in the detection of ten new triple candidates with outer periods shorter than $\sim$1300$\,$d. For five of them, we were able to constrain the outer orbit by analyzing independently the short-cadence (SC) and FFI data and to derive the minimum mass of the third body with a precision better than 4 per cent for SC and \tbf{11} per cent for FFI data. We then compared the results obtained from the two datasets and found that using the FFI data leads to \begin{enumerate*}[label=(\arabic*)]
\item a degradation of both the accuracy and precision of the tertiary mass determination for the tightest EBs and
\item an overall underestimation of the third component's mass. 
\end{enumerate*}%
However, we stress that our main conclusions on the nature of the detected signals do not depend on which dataset is used. This confirms the great potential of \TESS{} FFIs, which will allow us to search for rare objects such as substellar circumbinary companions and compact triple stellar systems.
\end{abstract}



\section{Introduction}\label{sec:intro}

Since its launch in 2018, the \textit{Transiting Exoplanet Survey Satellite} (\TESS{}; \citealt{2015JATIS...1a4003R}) has been surveying the majority of the sky, resulting in the detection of thousands of exoplanet candidates \citep{2021ApJS..254...39G}, solar-like oscillators \citep{2023A&A...669A..67H}, and eclipsing binaries (EBs; \citealt{2022ApJS..258...16P}) among $\sim$200,000 pre-selected stars observed in 2-min short-cadence (SC) mode during the 2-yr prime mission (Sectors 1--26). In addition, \TESS{} acquired images of each observing sector, referred to as full frame images, or FFIs, every 30 minutes. Over a billion stars were observed in this mode, therefore providing a huge data-mining archive for large ensemble studies, such as galactic archeology \citep{2020ApJ...889L..34S}. During the first extended mission (Sectors 27--55), the FFI cadence was reduced to 10 minutes, and a new 20-s ultra-short-cadence (USC) mode was implemented. \TESS{} is currently in its second extended mission (Sectors 56--69), and observes 2000 stars at 20-s cadence and 12,000 stars at 2-min cadence per sector, while FFIs are now retrieved at a shorter 200-s cadence.\\


Based on the SC data collected during the prime mission, \cite{2022ApJS..258...16P} produced a catalog\footnote{The Villanova \TESS{} EB catalog is available at \url{http://tessebs.villanova.edu/}} of 4584 EBs detected among the $\sim$200,000 pre-selected \TESS{} targets. EBs are among the most fundamental calibrators of stellar physics. In particular, it is possible to measure the masses and radii of each component of a double-lined (SB2) eclipsing binary with an exquisite precision, better than $\sim$1--3 per cent, and to derive its age without resorting to more advanced stellar modeling \citep{2021MNRAS.508.5687H}. For example, EBs hosting oscillating red giants are valuable benchmarks to test the ability of ensemble asteroseismology to reproduce the stellar properties \citep{2021A&A...648A.113B} and to calibrate model parameters, such as the mixing-length parameter \citep{2018MNRAS.475..981L}. EBs have also proven to be powerful tools for distance determinations of Local Group galaxies \citep{2019Natur.567..200P}. Finally, EBs that belong to multiple stellar systems are ideal to study the orbital architecture and dynamics \citep{2020MNRAS.499.3019M,2023MNRAS.521.1908M}.\\

Stellar multiplicity is an ubiquitous outcome of star formation. Indeed, almost half of solar-type stars are in binaries or multiple systems ($\sim$35 and 10 per cent of the total population, respectively), as reported by \cite{2010ApJS..190....1R} from a survey of stellar multiplicity in the solar neighborhood. The occurrence rate of close binaries with tertiary companions ($\sim$20 per cent of the EB population) was confirmed by \cite{2014AJ....147...45C}, who analyzed a sample of 1279 \Kepler{} eclipsing binaries using the so-called eclipse timing variation method. Eclipse timing variations (ETVs) are generally attributed to the light-travel time effect (LTTE; \citealt{1990BAICz..41..231M}), also known as the R\o{}mer delay, caused by the presence of a circumbinary third body. The ETV method is therefore particularly suitable for identifying compact hierarchical triples (CHTs; \citealt{2022Galax..10....9B}) and substellar circumbinary companions \citep{2016A&A...587A..82W,2018A&A...620A..72W,2021A&A...647A..65W} from both space-based and ground-based surveys. Increasing the detection of these objects will considerably improve our knowledge of the formation of stars and planets in binaries or multiple stellar systems \citep{2019Galax...7...84M,2021Univ....7..352T}. On the one hand, there is a clear deficit of CHTs with binary periods $\lesssim\,$1$\,$d and outer periods $\lesssim\,$200$\,$d \citep{2016MNRAS.455.4136B}. 
\tbf{On the other hand, only 15--20 circumbinary planets are presently known \citep{2023Univ....9..455K}, with the majority having been discovered through their transits (see \eg \citealt{2021AJ....162..234K}) or ETVs (see \eg \citealt{2017MNRAS.468.2932G}).}\\

The \TESS{} FFIs contain hundreds of thousands of EBs \citep{2021tsc2.confE.163K}, among which several thousand are expected to be located in the continuous viewing zones (CVZs). As they benefit from long-duration observations, CVZ EBs are promising targets for finding CHTs and circumbinary planets using the ETV method. \tbf{This was recently demonstrated by \cite{2024AA...685A..43M}, who reported the detection of 125 new triple candidates in the \TESS{} northern CVZ. The \TESS{} mission has also enabled the discovery of a steadily growing number of new triply eclipsing triple systems. Such systems are particularly valuable as their stellar and orbital properties can be fully determined from a joint photodynamical analysis of the photometric
light curves, ETVs, and spectral energy distributions, coupled with radial velocities and stellar evolutionary tracks. Some 30 systems were analyzed in this way using \TESS{} photometric data \citep{2020MNRAS.496.4624B,2022MNRAS.510.1352B,2020MNRAS.498.6034M,2022MNRAS.513.4341R,2023MNRAS.521..558R,2024A&A...686A..27R,2023MNRAS.526.2830C}.} Given the expected large number of CVZ EBs to be detected in the FFIs, it is necessary to use a fast, automated method for performing precise eclipse timing. In this paper, the first in a series, we aim to validate the light-curve extraction from the \TESS{} FFIs using the calibrated 2-min cadence light curves, and to demonstrate the potential of our method for recovering the ETV signal of a third body using the FFIs. The paper is organized as follows. We describe our target selection in Section~\ref{sec:data}, as well as the \TESS{} observations and data reduction. In Section~\ref{sec:method}, we present our method to measure precise eclipse times and to determine third-body solutions from ETVs. In Section~\ref{sec:discussion}, we discuss the results obtained for five EBs with a firm third-body detection, focusing on the comparison \tbf{between different approaches and} between SC and FFI results. Finally, in Section~\ref{sec:summary}, we summarize the conclusions of this work and outline future prospects.

\section{Targets and data}\label{sec:data}

\subsection{Target selection}

To select \TESS{} EB targets suitable for long-duration ETV surveys from both the SC and FFI datasets, we made use of the Villanova \TESS{} EB catalog \citep{2022ApJS..258...16P}, which lists 4584 EB targets observed in SC mode during the prime mission. Among them, we identified $\sim$400 targets in the CVZs, that is, with an ecliptic latitude $|\beta| > 78^\circ$. As illustrated in Figure~2 of \cite{2022ApJS..258...16P}, the catalog contains a variety of systems, from contact and ellipsoidal binaries with periods as short as 0.1$\,$d to wide detached binaries with periods of tens of days. For this reason, \cite{2022ApJS..258...16P} adopted the morphology coefficient defined by \cite{2012AJ....143..123M} to describe the ``detachedness'' of a binary. This parameter continuously varies between 0 and 1, where 0 corresponds to the widest detached binaries and 1 to contact binaries and ellipsoidal variables. The timing method described in this paper was initially developed for the ETV analysis of detached and semi-detached EBs, that is, EBs with a morphology coefficient less than $\sim$0.7 \citep{2020MNRAS.499.3019M,2024MNRAS.527...53M}.\\ 

As a part of the selection process, we applied our timing procedure to the full set of $\sim$400 CVZ targets previously identified in the Villanova \TESS{} EB catalog. We also performed a visual inspection of the light curves to identify detached and semi-detached EBs in the sample. 
Thus, from our original sample, we were able to measure the eclipse times of $\sim$100 detached and semi-detached EBs with a period shorter than 27$\,$d, \ie the observing duration of one \TESS{} sector. Finally, we inspected the resulting $O-C$ (observed minus calculated times) curves of these $\sim$100 EBs, and we found that 10 of them show ETV signals with periods less than $\sim$1300$\,$d (corresponding to the maximum time span of the observations used in this work). It is worth noting that the third-body occurrence rate in our sample, $\sim$10 per cent, is consistent with that found by \cite{2014AJ....147...45C} for a range of outer periods up to $\sim$1400$\,$d. For five of our ten newly identified triple systems, the good coverage of the outer orbit with observations in SC mode allowed us to make a direct comparison with the results obtained using the FFI data. These five systems were observed in Sectors 14--26, 40--41, and 47--60, implying a coverage of their outer orbits of at least 60 per cent. They are listed in Table~\ref{tab:tic}, \tbf{along with the parallaxes and renormalized unit weight errors (RUWEs; \citealt{LL:LL-124}) from the \Gaia{} Data Release 3 (DR3) catalog \citep{2022yCat.1355....0G}. All of our systems have a RUWE greater than 1.4, except for \ticfour{}, for which no astrometric solution is available. This tends to confirm the presence of a tertiary companion, as noted by \cite{2021ApJ...907L..33S}.}

\begin{table}
    \centering
    \begin{minipage}{180mm}
        \caption{Properties of the detected systems taken from the Villanova \TESS{} EB catalog. The five systems investigated in this work are marked by an asterisk.}
        \label{tab:tic}
        {
        \renewcommand{\arraystretch}{1.0}
        \begin{tabular}{@{}lcccccc@{}}
            \hline
            \hline
            TIC           & $T_0$               & $P_1$              & Morph. & $T_{\rm mag}$ & \tbf{Parallax}$^a$         & \tbf{RUWE}$^a$ \\
                          & BJD$-245\,7000$     & (d)                &        &               & \tbf{(mas)}                &                \\
            \hline
            38699825      & $1410.972\,4(40)$   & $2.084\,644(11)$   & 0.592  & 8.296         & $\mbf{3.7003 \pm 0.0319}$  & \tbf{2.416}    \\
            141685465     & $1416.199\,5(14)$   & $7.807\,423(97)$   & 0.220  & 11.3742       & $\mbf{2.7643 \pm 0.0254}$  & \tbf{1.882}    \\
            150361911     & $1545.543\,040(99)$ & $2.493\,8523(28)$  & 0.461  & 12.1631       & $\mbf{1.1225 \pm 0.0196}$  & \tbf{1.937}    \\
            219900027*    & $1683.750\,363(15)$ & $0.515\,26471(79)$ & 0.684  & 11.2024       & $\mbf{1.7667 \pm 0.0394}$  & \tbf{2.092}    \\
            229771234*    & $1931.061\,6(66)$   & $0.820\,95849(30)$ & 0.722  & 10.7665       & $\mbf{14.0558 \pm 0.1461}$ & \tbf{15.674}   \\
            259006185*    & $1901.860\,148(45)$ & $1.939\,6542(16)$  & 0.346  & 9.11642       & $\mbf{5.2480 \pm 0.0166}$  & \tbf{1.468}    \\
            278956474$^b$ & $1327.960\,81(26)$  & $5.488\,050(12)$   & 0.314  & 12.9637       & $\mbf{1.0165 \pm 0.0354}$  & \tbf{4.230}    \\
            356896561*    & $1685.409\,425(11)$ & $2.253\,35087(11)$ & 0.455  & 9.1494        & $\mbf{4.84 \pm 1.66}$      & \tbf{--}       \\
            373915220     & $1329.027\,38(34)$  & $6.470\,6807(19)$  & 0.149  & 11.4541       & $\mbf{2.0332 \pm 0.0307}$  & \tbf{1.897}    \\
            424461577*    & $1712.102\,37(17)$  & $0.744\,82334(12)$ & 0.654  & 10.1535       & $\mbf{2.3863 \pm 0.0339}$  & \tbf{2.458}    \\
            \hline 
        \end{tabular}
        }
    \end{minipage}
    \tablecomments{$^a$ \tbf{The parallax and RUWE values are from the \Gaia{} DR3 catalog. For \ticfour, we adopted the parallax from \cite{1997AA...323L..49P}.}\\
    $^b$ \ticsix{} was identified as a doubly eclipsing quadruple system by \cite{2020AJ....160...76R}.
    }
\end{table}

\subsection{Short-cadence data}

We first analyzed the SC data collected by \TESS{} for our initial sample of $\sim$400 CVZ targets. These data were generated by the \TESS{} Science Processing Operations Center (SPOC; \citealt{2016SPIE.9913E..3EJ}) and made available on the Mikulski Archive for Space Telescopes (MAST)\footnote{\url{https://archive.stsci.edu/}}. We employed the \astroquery{} package \citep{2019AJ....157...98G} to download the SC light curves of our targets from Sectors 1 to 60.
In our analysis, we chose to use the Pre-search Data Conditioning Simple Aperture Photometry (PDCSAP) light curves \citep{2012PASP..124.1000S,2012PASP..124..985S,2014PASP..126..100S}, which have been corrected for instrumental effects and contamination by nearby stars. Then, we normalized the light curves by dividing by the median flux value of each sector.

\subsection{Full-frame image data}

\begin{figure}
    \centering
    \includegraphics[trim = 2.9cm 4.0cm 2.4cm 2.8cm,clip,width=0.5\columnwidth,angle=0]{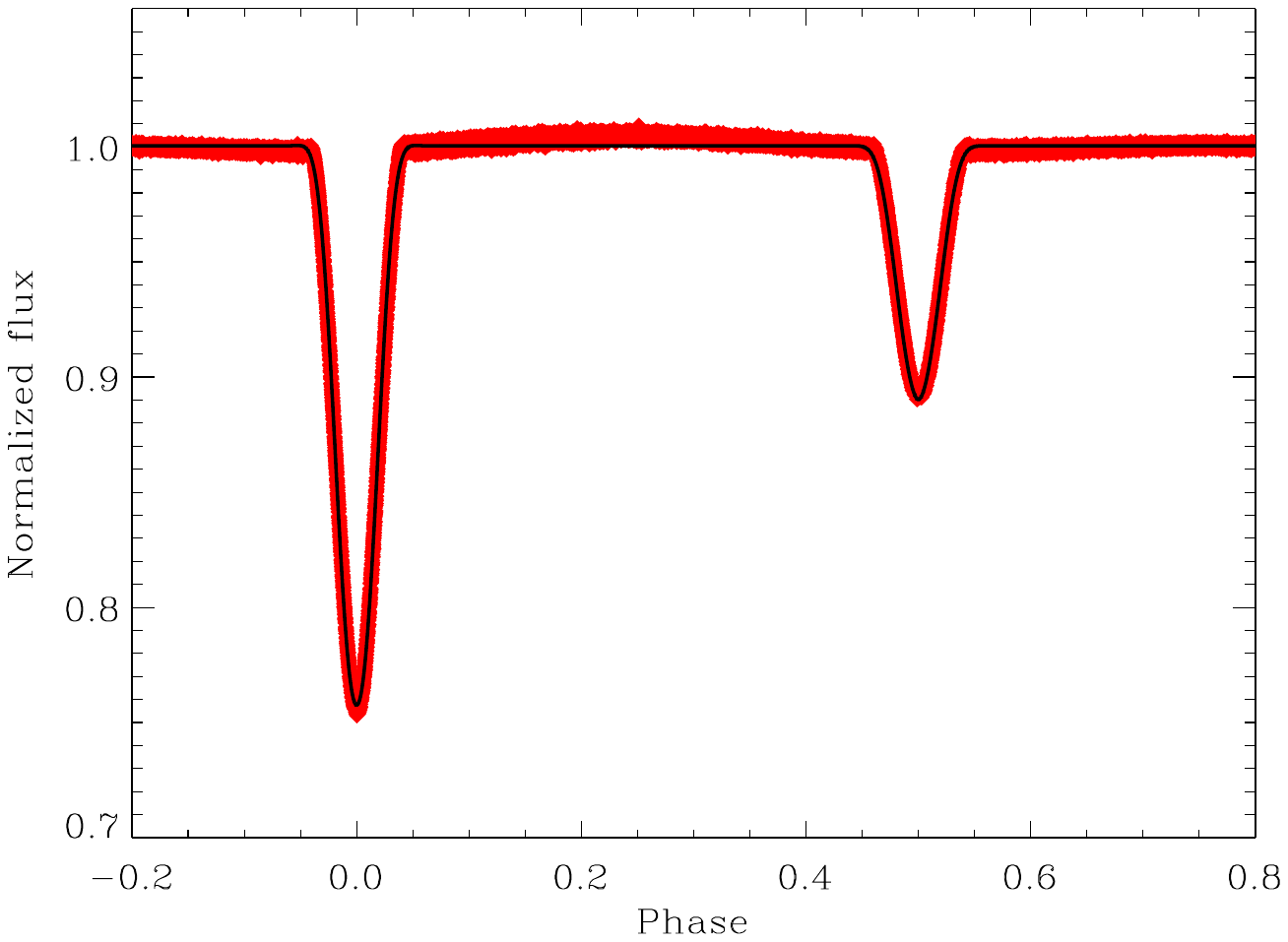}
    \includegraphics[trim = 2.9cm 2.4cm 2.4cm 2.8cm,clip,width=0.5\columnwidth,angle=0]{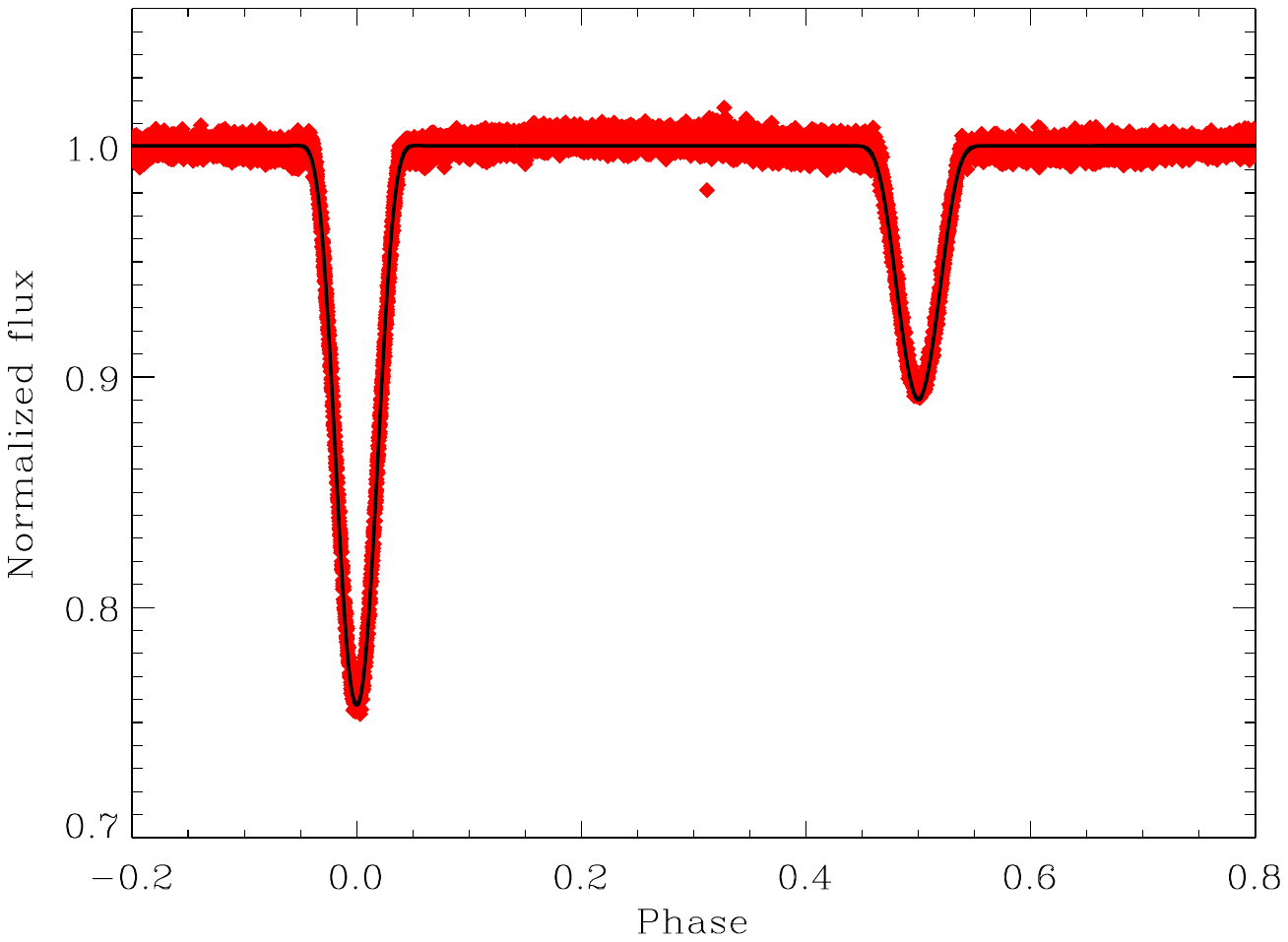}
    \caption{The \TESS{} SC (top) and FFI (bottom) light curves of \ticfour{} (presented here is Sector~57). Red symbols denote the observations while the solid line corresponds to the best-fit model of the SC light curve determined using the procedure described in Section~\ref{sec:times}.}
    \label{fig:LC}
\end{figure}

For all of our CVZ targets, we also extracted FFI light curves from all available sectors\footnote{The FFI data for Sectors 58–60 and later cannot be accessed via \eleanor{}.} using the \eleanor{} package \citep{2019PASP..131i4502F} with the default settings. We briefly describe the main steps of the light-curve extraction with \eleanor{}. First, we download a $31 \times 31$ pixel postcard of the FFIs centered on each target using the \tesscut{} tool \citep{2019ascl.soft05007B} called by \eleanor{}. A $13 \times 13$ target pixel file (TPF) is then extracted from each postcard, and aperture photometry is performed on the TPF. We used a fixed aperture of $3 \times 3$ pixels from the \eleanor{} default library, which yields the lowest point-to-point scatter for the range of magnitudes of our targets \citep{2020MNRAS.498.5972N}. \Eleanor{} provides three different types of light curves from aperture photometry: raw light curves, corrected light curves, and principal component analysis (PCA) light curves. For our analysis, we used the corrected light curves, where signals correlated to pixel position, measured background and time have been removed, and we excluded all flux measurements with nonzero quality flags. Finally, the FFI light curves were normalized in the same way as the SC light curves. The SC and FFI light curves of the five investigated targets are shown in Figure~\ref{fig:LC} and Appendix~\ref{sec:LC}.

\section{Method and analysis}\label{sec:method}

\subsection{Bayesian inference method}\label{sec:bayes}

Recently, \cite{2024MNRAS.527...53M} published the first results of the \Solaris{}\footnote{\url{http://projektsolaris.pl/en/homepage/}} ground-based photometric survey, which led to the detection of a new CHT candidate, GSC 08814-01026, using the ETV method. The present section aims to describe the Bayesian formalism adopted in this work, as well as in \cite{2024MNRAS.527...53M}, for the derivation of the times of minima.\\

According to Bayes' theorem, the posterior probability density of the parameters $\Theta$ given data $D$ is stated as
\begin{equation}
    P(\Theta | D) = \frac{ P(\Theta) \, P(D | \Theta)}{P(D)},\label{eq:Bayes}
\end{equation}
where $P(\Theta)$ is the prior probability density of the parameters, $P(D)$ is the global normalization likelihood and $P(D | \Theta)$ is the likelihood ${\cal L}$ of the data given the parameters. Here, we define the likelihood ${\cal L} = \exp(-\chi^2/2)$ with
\begin{equation}
    \chi^2 = \sum^{N}_{i=1} \, \left(\frac{D_i-M_i(\Theta)}{\sigma_i} \right)^2,\label{eq:chi2}
\end{equation}
where $N$ denotes the number of observed data points, $\sigma_i$ refers to the associated uncertainties, and $M(\Theta)$ is the model function used to fit the data (see Sections~\ref{sec:times} and~\ref{sec:etv}). The sampling of the posterior probabilities can be done by exploring the parameter space using Markov Chain Monte Carlo (MCMC) methods, such as the Metropolis--Hastings algorithm (MH; \citealt{1953JChPh..21.1087M,10.1093/biomet/57.1.97}). Our implementation of the MH algorithm works as follows. We set a chain of N points with starting points $\Theta^{t}$ taken randomly from appropriate distributions. The new set of parameters $\Theta^{t'}$ is computed using a random walk as
\begin{equation}
    \Theta^{t'} = \Theta^{t} + \alpha_{\rm rate} \, \Delta \Theta,
\end{equation}
where $\Delta \Theta$ is given by a multivariate normal distribution with independent parameters and $\alpha_{\rm rate}$ is an adjustable parameter that is reduced by a factor of two until the rate of acceptance of the new set exceeds 25 per cent, marking the end of the burn-in phase (first 10 per cent of the chain). We assume here that either set has the same probability, \ie $P(\Theta^{t})=P(\Theta^{t'})$. From Equation~(\ref{eq:Bayes}), we can then define the ratio
\begin{equation}
    r=\frac{P(\Theta^{t'} | D)}{P(\Theta^{t} | D)}=\frac{P(D | \Theta^{t'})}{P(D | \Theta^{t})},
\end{equation}
which is equivalent to the ratio of the likelihoods, known as the Bayes factor. The acceptance probability of the new set, $\Theta^{t'}$, is computed as
\begin{equation}
    \alpha(\Theta^{t},\Theta^{t'}) = {\rm min}\,(1,r) = {\rm min}\left(1,\frac{P(D | \Theta^{t'})}{P(D | \Theta^{t})}\right).
\end{equation}
The new values of the parameters are accepted if $\beta \leq \alpha$, where $\beta$ is a random number drawn from a uniform distribution over the interval $[0,1]$, and rejected otherwise. We derive the posterior probability distribution of each parameter from the Markov chain after rejecting the initial burn-in phase. 
For all parameters, we compute the median and the credible intervals at 16 per cent and 84 per cent, corresponding to a 1$\sigma$ interval for a normal distribution. The advantage of this percentile definition over the mode (maximum of the posterior distribution) or the mean (average of the distribution) is that it is conservative with respect to any change of variable over these parameters (see Section~\ref{sec:comparison}).

\subsection{Determination of times of minima}\label{sec:times}

In order to measure the times of eclipse minima, 
we approximated the phased light curves of our targets using the phenomenological model of \citet{2015A&A...584A...8M}, which analytically describes the eclipse profile as
\begin{equation}
    f(t_i,\Theta) = \alpha_0 + \sum^{n_{\rm e}}_{k=1} \alpha_k \, \psi(t_i,T_k,d_k,\Gamma_k,C_k),\label{eq:ecl_profile} 
\end{equation}
where $\alpha_0$ is the flux zero-point shift, $\alpha_k$ is a scaling coefficient of the eclipse profile function $\psi$, and $n_{\rm e}$ is the number of eclipses during one cycle. Among the seven different model functions proposed by \citet{2015A&A...584A...8M}, we chose the form:
\begin{equation}
    \psi(t_i,T,d,\Gamma,C) = \Bigg\{ 1 + C \bigg( \frac{t_i-T}{d} \bigg)^2 \Bigg\} \Bigg\{ 1-\bigg\{ 1-\exp \bigg[ 1-\cosh \bigg( \frac{t_i-T}{d} \bigg) \bigg] \bigg\}^\Gamma \Bigg\},
\end{equation}
which is suitable for detached and semi-detached EBs (see \citealt{2020MNRAS.499.3019M} and  \citealt{2024MNRAS.527...53M} who used a simpler form of $\psi$). Here, $T$, $d$, $\Gamma$, and $C$ are the time of minimum, the eclipse width, the kurtosis, and the scaling parameter, respectively. For each individual eclipse, the time of minimum is estimated as
\begin{equation}\label{eq:ecl_mod}
    \begin{split}
        T_k & = T_{0,k} + P E + \Delta_k \\
        & = T_{0,k} + P \times {\rm round} \bigg( \frac{t_i-T_{0,k}}{P} \bigg) + \Delta_k,
    \end{split}
\end{equation}
where $T_{0,k}$, $P$, $E$, and $\Delta_k$ are the reference time of eclipse, the orbital period, the epoch, and the $O-C$ time difference, respectively. \tbf{Additionally, the out-of-eclipse variability can be taken into account by including the contribution of the O'Connell and proximity effects in Equation~(\ref{eq:ecl_profile}). The O'Connell effect refers to the height difference between successive out-of-eclipse maxima that can be seen in the light curves of some eclipsing binaries \citep{1951PRCO....2...85O,1968AJ.....73..708M}. This asymmetry may be caused by chromospheric spots, by the capture of circumstellar material or by a hot spot formed by the impact of a mass-transferring gas stream \citep{2003ChJAA...3..142L,2009SASS...28..107W}. As noted by \cite{2009SASS...28..107W}, the O'Connell effect contribution is well approximated by a single sine function}
\boldm{
\begin{equation}
    f_{\rm c}(t_i,\Theta) =  A_{\rm c} \sin \Bigg[ 2 \pi \bigg( \frac{t_i-T_{k=1}}{P} \bigg) \Bigg],\label{eq:oconnell} 
\end{equation}
}%
\tbf{where \boldm{$A_{\rm c}$} is the amplitude of the O'Connell effect and \boldm{$T_{k=1}$} represents the times of primary minima, while the contribution of proximity effects (deformation of the components, gravity darkening and mutual irradiation) can be described by a linear combination of \boldm{$n_p$} cosine functions as}
\boldm{
\begin{equation}
    f_{\rm p}(t_i,\Theta) = \sum^{n_{\rm p}}_{n=1} A_{{\rm p},n} \cos \Bigg[ 2 \pi n \bigg( \frac{t_i-T_{k=1}}{P} \bigg) \Bigg],\label{eq:proximity} 
\end{equation}
}%
\tbf{where the \boldm{$A_{{\rm p},n}$} terms are the coefficients of the cosine series. Thus, the full light-curve model is based on the set of parameters}
$\Theta =$ ($\alpha_0,\alpha_k,T_{0,k},d_k,\Gamma_k,C_k,\Delta_k,P$,\boldm{$A_{{\rm p},n}$},\boldm{$A_{\rm c}$}), where $\Delta_k$ is a vector with size equal to the number of cycles.\\

In the first step of the modeling process, we applied our MCMC procedure to the PDCSAP light curve of each target assuming that the eclipses are strictly periodic (i.e., $\Delta_k = 0$ for each eclipse). \tbf{We accounted for the contribution of the O'Connell and proximity effects observed in the light curves of \ticone{} and \ticfive{} by fitting the amplitudes defined in Equations~(\ref{eq:oconnell}) and~(\ref{eq:proximity}). Here, we limited the series up to fourth order (\boldm{$n_{\rm p} = 4$}).} For each target, we generated ten chains of $10^5$ points each as described in Section~\ref{sec:bayes}. From each chain, we obtained a set of parameters that best fits the light curve of a single sector (randomly selected among the available sectors). We then computed the chi-squared ($\chi^2$) for all sets of parameters as defined in Equation~(\ref{eq:chi2}) and adopted as reference the set with the smaller value of $\chi^2/N$ among the ten MCMC chains. In Figure~\ref{fig:LC} and Appendix~\ref{sec:LC}, we compare our best-fit models, derived from a single sector, to the SC and FFI light curves of the five investigated systems. 
In the second step, we used our MCMC algorithm to determine the times of minima from both the PDCSAP and FFI light curves. Here, we generated a chain of $10^4$ points for each available SC sector. During the fitting, we left $\Delta_k$ as a free parameter and we fixed the other parameters to their best-fit values obtained in the initial fit. We then calculated the times of minima from the best-fit values of the parameter $\Delta_k$ using Equation~(\ref{eq:ecl_mod}). We repeated this step for all FFI sectors. Finally, the measured times of all eclipses occurring during data gaps were removed, an iterative sigma-clipping technique was applied to the remaining eclipse times, and the error bars were properly rescaled (see Section~\ref{sec:etv}). For the five systems studied here, we provide the times of minima derived from both the PDCSAP and FFI light curves in Table~\ref{tab:ecl_times}.

\begin{table}
    \centering
    \begin{minipage}{155mm}
        \caption{Measured times of minima for the systems investigated in this work.}
        \label{tab:ecl_times}
        {
        \renewcommand{\arraystretch}{1.0}
        \begin{tabular}{@{}lccccrc@{}}
            \hline
            \hline
            TIC & Time & Cycle & 1$\sigma$ error & $\Delta_{\rm obs}$ & \multicolumn{1}{c}{${\rm O}-{\rm C}$} & Obs.\ mode \\
            & BJD$-245\,7000$ & no. & (d) & (s) & \multicolumn{1}{c}{(s)} & \\
            \hline
            356896561 & $1684.282\,68$   &  $-$532.5    &    $0.000\,27$   & 61.7 &    29.0 & SC \\
            356896561 & $1685.408\,82$   &  $-$532.0    &    $0.000\,13$   & 14.9 & $-$18.3 & SC \\
            356896561 & $1686.535\,95$   &  $-$531.5    &    $0.000\,30$   & 53.7 &    19.9 & SC \\
            356896561 & $1687.662\,17$   &  $-$531.0    &    $0.000\,13$   & 14.2 & $-$20.2 & SC \\
            356896561 & $1688.789\,19$   &  $-$530.5    &    $0.000\,26$   & 43.3 &    8.4  & SC \\
            356896561 & $1689.915\,52$   &  $-$530.0    &    $0.000\,13$   & 13.4 & $-$22.1 & SC \\
            356896561 & $1691.042\,49$   &  $-$529.5    &    $0.000\,27$   & 38.1 &    2.1  & SC \\
            356896561 & $1697.802\,65$   &  $-$526.5    &    $0.000\,27$   & 45.5 &    6.4  & SC \\
            356896561 & $1698.929\,08$   &  $-$526.0    &    $0.000\,13$   & 23.9 & $-$15.7 & SC \\
            356896561 & $1700.056\,12$   &  $-$525.5    &    $0.000\,29$   & 54.5 &    14.4 & SC \\
            356896561 & $1701.182\,45$   &  $-$525.0    &    $0.000\,13$   & 24.7 & $-$15.9 & SC \\
            \hline
        \end{tabular}
        }
    \end{minipage}
    \tablecomments{Half-integer cycle numbers refer to secondary eclipses. $O-C$ refers to the ETV residuals. The uncertainties were determined as described in Section~\ref{sec:etv}. \\
    (This table is available in machine-readable form.)}
\end{table}

\subsection{Eclipse timing variation analysis}\label{sec:etv}

\begin{figure}
    \centering
    \includegraphics[trim = 2.0cm 3.7cm 3.5cm 1.5cm,clip,width=0.6\columnwidth,angle=0]{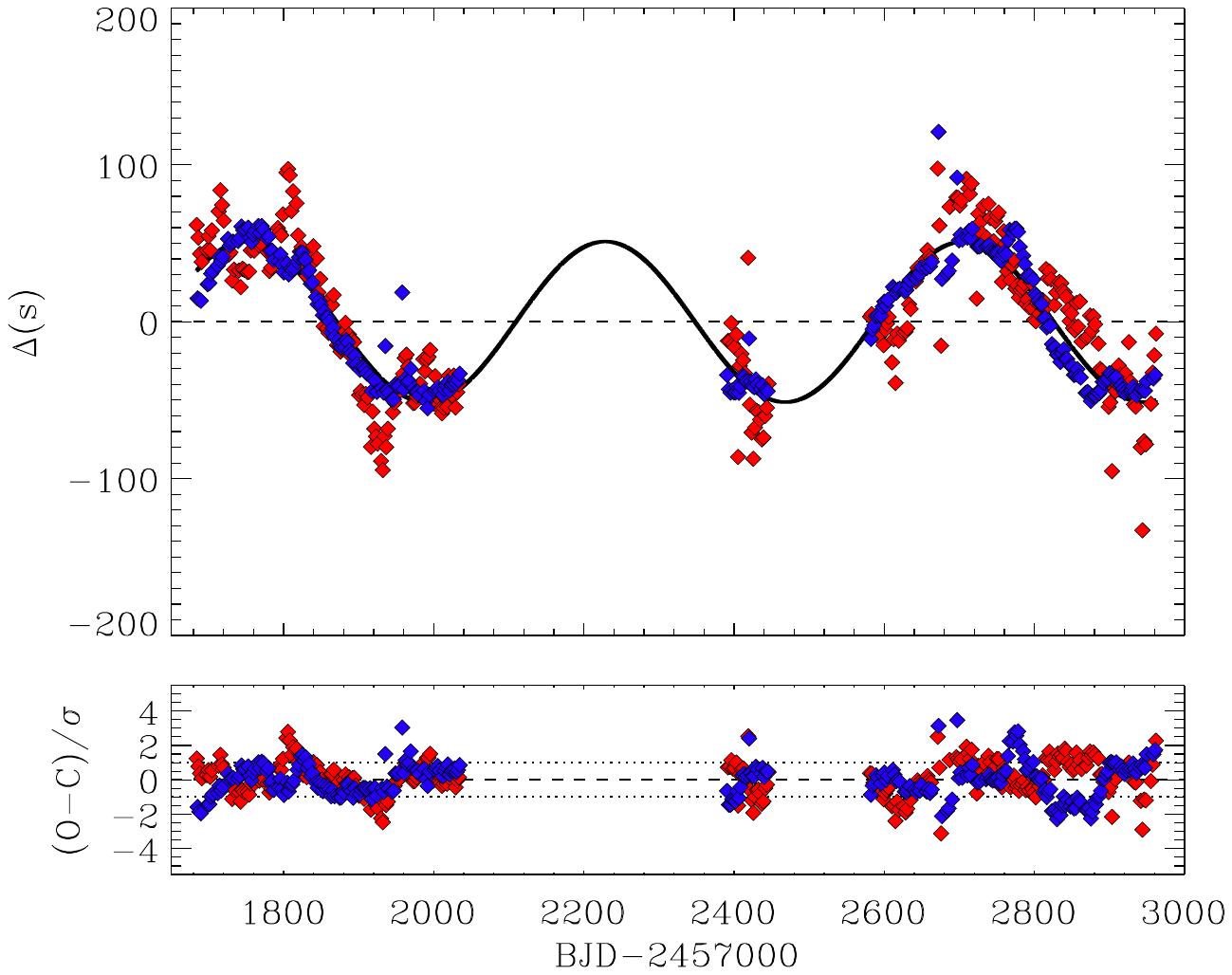}
    \includegraphics[trim = 2.0cm 2.0cm 3.5cm 1.5cm,clip,width=0.6\columnwidth,angle=0]{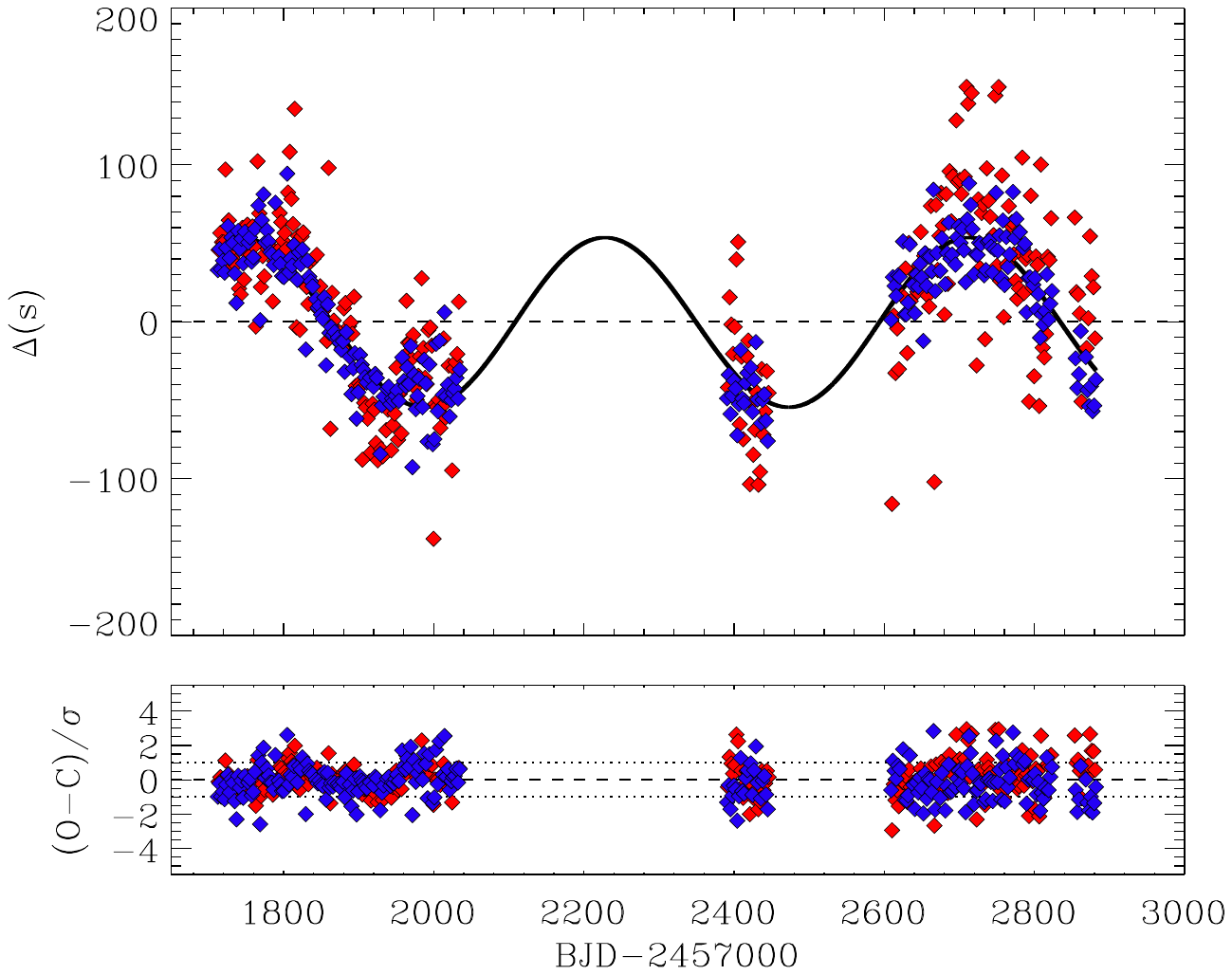}
    \caption{ETVs for \ticfour{} determined from the SC (top) and FFI (bottom) light curves. Blue and red diamonds refer to the primary and secondary eclipse times, respectively, derived as explained in Section~\ref{sec:times}. The best-fit ETV solution is overplotted as a solid line. Fitting residuals are shown in lower panels.}
    \label{fig:ETV_curve}
\end{figure}

A common method to search for ETVs is to construct an $O-C$ diagram showing the difference between the observed and calculated times of minima, \ie
\begin{equation}
    \Delta_{\rm obs} = T_{\rm o}(E) - T_{\rm c}(E) = T_{\rm o}(E) - T_0 - P E,
\end{equation}
where $T_{\rm o}(E)$ and $T_{\rm c}(E)$ refer to the observed and calculated times of minima at epoch $E$, respectively. Here, the values of $T_0$ and $P$ are taken from the best-fit light-curve models obtained in Section~\ref{sec:times}. The resulting $O-C$ diagrams of the five systems investigated in this work are shown in Figure~\ref{fig:ETV_curve} and Appendix~\ref{sec:ETV_curve}.\\

As mentioned previously, ETVs can be attributed to the LTTE caused by the gravitational influence of a third body orbiting an EB. There are, however, various other mechanisms capable of inducing ETVs in an EB, such as the apsidal motion and dynamical perturbation effects (see \eg \citealt{2015MNRAS.448..946B}, for a review). Apsidal motion consists in the rotation of the line of apsides in an eccentric binary and has an  anticorrelated effect for primary and secondary minima variations. In the following, apsidal motion is precluded as the five EBs presented here are nearly circular. Furthermore, in the presence of a third body, the EB undergoes a number of dynamical perturbations, implying that its orbit is no longer Keplerian. The dynamical perturbations affect all the orbital elements of the EB and, thus, the occurrence times of the eclipses. \cite{2016MNRAS.455.4136B} showed that the  amplitude of the dynamical contribution for a circular EB orbit can be written as
\begin{equation}
{\mathcal A}_{\rm dyn} = \frac{1}{2\pi} \frac{M_{\rm B}}{M_{\rm A}+M_{\rm B}} \, \frac{P_1^2}{P_2} \big(1-e_2^2\big)^{-3/2},\label{eq:Adyn} 
\end{equation}
where $M_A$ and $M_B$ are the masses of the eclipsing pair and the third body, respectively, and $e_2$ is the eccentricity. The subscript `1' refers to the inner orbit, \ie the eclipsing pair A, while the subscript `2' refers to the relative orbit between the eclipsing pair A and the third body B, hereafter referred to as the AB system. Equation~(\ref{eq:Adyn}) attests that the dynamical ETV contribution is expected to be more important for the systems in our sample that have the smallest $P_2/P_1$ ratio (\ticthree{} and \ticfour{}). However, in the present study, we neglect the possible contribution of the dynamical perturbations.
We thus modeled the ETVs in the mathematical form of LTTE \citep{2015MNRAS.448..946B,2016MNRAS.455.4136B}:\\
\begin{equation}
    \Delta_{\rm mod} = c_0 + c_1 E - \frac{a_{\rm A}\sin i_2}{c} \, \frac{\big(1-e_2^2\big)\sin(\upsilon_2+\omega_2)}{1+e_2\cos \upsilon_2},\label{eq:OC_mod}
\end{equation}
where $c_0$ and $c_1$ are factors that correct the respective values of $T_0$ and $P_1$ for the ETV effect, $a_{\rm A}$ is the semi-major axis of the EB's barycentric orbit, $c$ is the speed of light, and $i_2$, $e_2$, $\upsilon_2$, and $\omega_2$ are the inclination, eccentricity, true anomaly, and argument of periastron of the third component's relative orbit, respectively. For simplicity, we use the semi-amplitude of the LTTE ETVs defined as \citep{1952ApJ...116..211I}
\begin{equation}
    A_{\rm LTTE} = \frac{a_{\rm A}\sin i_{\rm AB}}{c} = \frac{{\mathcal A}_{\rm LTTE}}{(1-e_2^2 \cos^2 \omega_2)^{1/2}}.\label{eq:Altte}
\end{equation} 
The true anomaly $\upsilon_2$ is given by 
\begin{equation}
    \tan\frac{\upsilon_2}{2} = \sqrt{\frac{1+e_2}{1-e_2}} \, \tan\frac{E_2}{2},
\end{equation}
where $E_2$ is the eccentric anomaly. This latter can be found using a fixed-point method to solve the Kepler's equation for any time $t$:
\begin{equation}
\frac{2\pi}{P_2}\,(t-T_2) = E_2 - e_2\,\sin E_2,
\end{equation}
where $P_2$ and $T_2$ are the orbital period and the time of periastron passage of the third component's relative orbit, respectively.
In summary, our ETV model consists of the parameters $\Theta = (c_0,c_1,
e_2,\omega_2,P_2,T_2,A_{\rm LTTE})$.\\

In this work, we aim to demonstrate the feasibility of accurately measuring the third-body parameters from the ETV analysis of FFI targets. Therefore, we processed the two sets of ETV curves, derived from the corresponding FFI and SC light curves,  independently. Another important aspect of our analysis is the determination of reliable uncertainties on the derived parameters. For this, we performed an initial fit of each ETV curve using a chain of $10^6$ points. Based on the best-fit preliminary solution, we applied an iterative 4$\sigma$-clipping technique to exclude possible outliers. In each iteration, we first computed the normalized chi-squared ($\chi^2/N$) of the best fit for both the primary and secondary ETV curves. Then, we rescaled the error bars of the primary and secondary eclipse times by multiplying them by $\sqrt{\chi^2_{\rm p}/N_{\rm p}}$ and $\sqrt{\chi^2_{\rm s}/N_{\rm s}}$, respectively. We removed the 4$\sigma$ outliers, and repeated the above steps until no more outliers were found. In order to obtain reliable uncertainties on the fitted parameters, we performed a new fit of the ETV curves using the rescaled error bars. Finally, we checked that the best-fit solution has $\chi^2_{\rm p}/N_{\rm p} \sim 1$ and $\chi^2_{\rm s}/N_{\rm s} \sim 1$ and that the mean error on primary and secondary eclipse times is of the same order of magnitude as the root mean square (rms) of the primary and secondary ETV residuals. The best-fit ETV solutions of our five targets are shown in Figure~\ref{fig:ETV_curve} and Appendix~\ref{sec:ETV_curve}. The values derived from the ETV fits of the SC data are given in Table~\ref{tab:ETV_param} and Appendix~\ref{sec:ETV_param}, together with their uncertainties.

\begin{table}
    \centering
    \begin{minipage}{130mm}
        \caption{Best-fit orbital parameters for \ticfour{} obtained from the SC data.}
        \label{tab:ETV_param}
        {
        \renewcommand{\arraystretch}{1.0}
        \begin{tabular}{@{}lccc@{}}
            \hline
            \hline
                                                                   &                 & 84 per cent      & 16 per cent      \\
            Parameter                                              & Median          & interval         & interval         \\
            \hline
            $P_{\rm A}$ (d)                                        & $2.253\,360405$ & $+0.000\,000047$ & $-0.000\,000046$ \\
            $T_0$ (BJD$-245\,7000$)                                & $1685.408\,645$ & $+0.000\,013$    & $-0.000\,010$    \\
            $A_{\rm LTTE}$ (s)                                     & 51.2            & +1.1             & $-$1.1           \\
            $K_{\rm A}$ (km\,s$^{-1}$)                             & 2.34            & +0.05            & $-$0.05          \\
            $P_{\rm AB}$ (d)                                       & 477.5           & +2.1             & $-$2.0           \\
            $T_{\rm AB}$ (BJD$-245\,7000$)                         & 2600.5          & +8.2             & $-$8.6           \\
            $e_{\rm AB}$                                           & 0.0054          & +0.0074          & $-$0.0040        \\
            $\omega_{\rm AB}$ ($^\circ$)                           & 10.3            & +6.2             & $-$6.5           \\
            $a_{\rm A}\sin i_{\rm AB}$ (au)                        & 0.1026          & +0.0022          & $-$0.0022        \\
            $f\!$($M_{\rm B}$) (M$_\odot$)                         & $0.000\,631$    & $+0.000\,042$    & $-0.000\,040$    \\
            $M_{\rm B} \, (i_{\rm AB}\!=90^\circ$) (M$_\odot$)     & 0.1425          & +0.0033          & $-$0.0033        \\
            $M_{\rm B} \, (i_{\rm AB}\!=90^\circ$) (M$_{\rm Jup}$) & 149.3           & +3.4             & $-$3.5           \\
            \boldm{$a_{\rm AB}$} \tbf{(mas)}                       & \tbf{7.5}       & \tbf{+2.5}       & \tbf{$-$2.5}     \\
            ${\mathcal A}_{\rm dyn}/{\mathcal A}_{\rm LTTE}$       & 0.1901          & +0.0014          & $-$0.0014        \\
            \hline 
        \end{tabular}
        }
    \end{minipage}
    \tablecomments{The values of $T_0$ and $P_{\rm A}$ were corrected by $c_0$ and $c_1$, respectively. The mass of the third component, $M_{\rm B}$, \tbf{the semi-major axis of the outer relative orbit,} \boldm{$a_{\rm AB}$}, and the dynamical amplitude, ${\mathcal A}_{\rm dyn}$, were computed assuming $i_{\rm AB}\!=90^\circ$ and $M_{\rm Aa} = M_{\rm Ab} = 1.00 \pm 0.01 \, {\rm M}_\odot$ (see the text).}
\end{table}

\section{Discussion}\label{sec:discussion}

\subsection{\tbf{Comparison with the \cite{1956BAN....12..327K} method}}

\begin{figure}
    \centering
    \includegraphics[trim = 2.6cm 2.4cm 2.4cm 1.5cm, clip,width=0.48\linewidth]{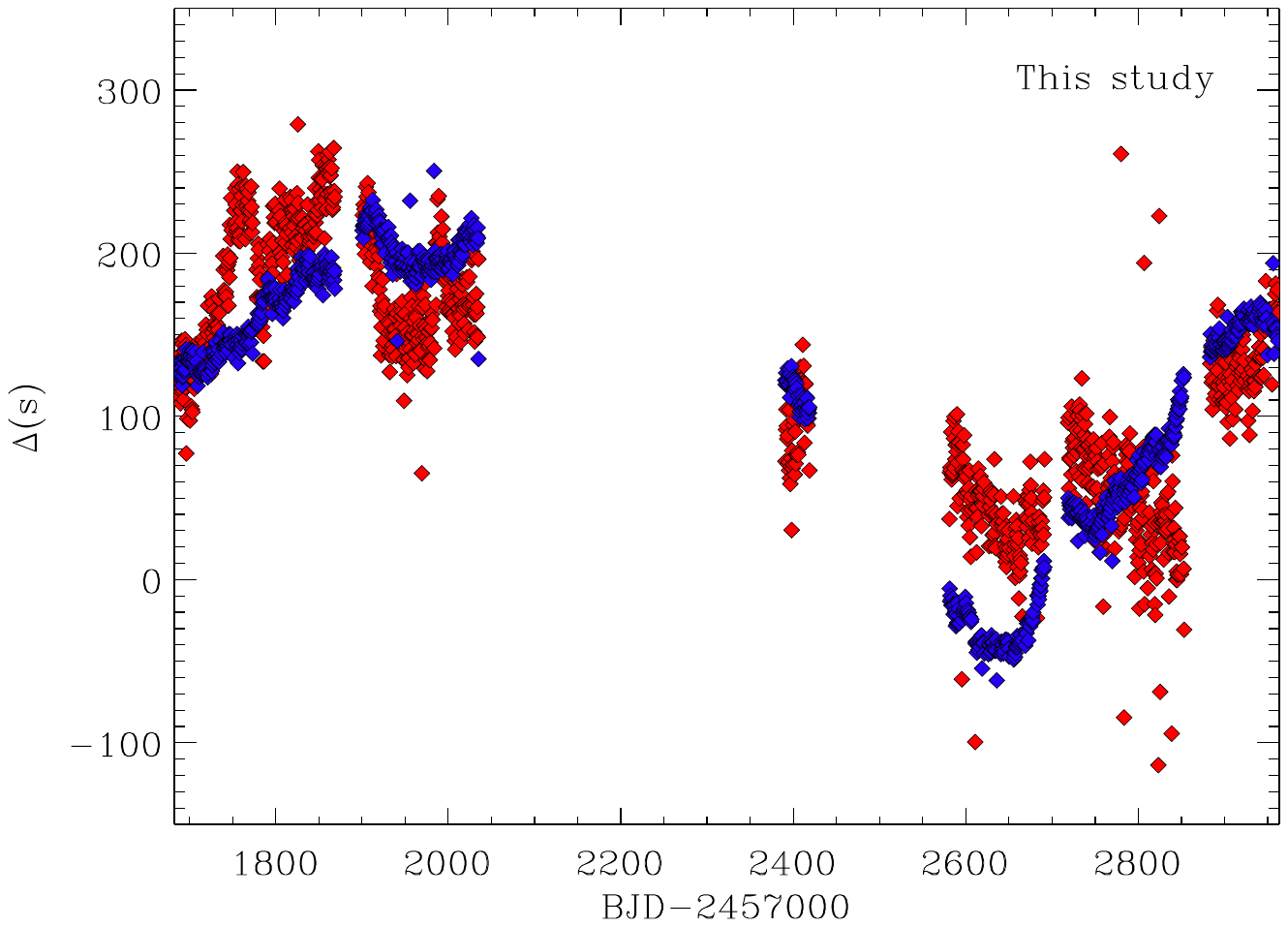}\hspace{0.02\textwidth}%
    \includegraphics[trim = 2.6cm 2.4cm 2.4cm 1.5cm, clip,width=0.48\linewidth]{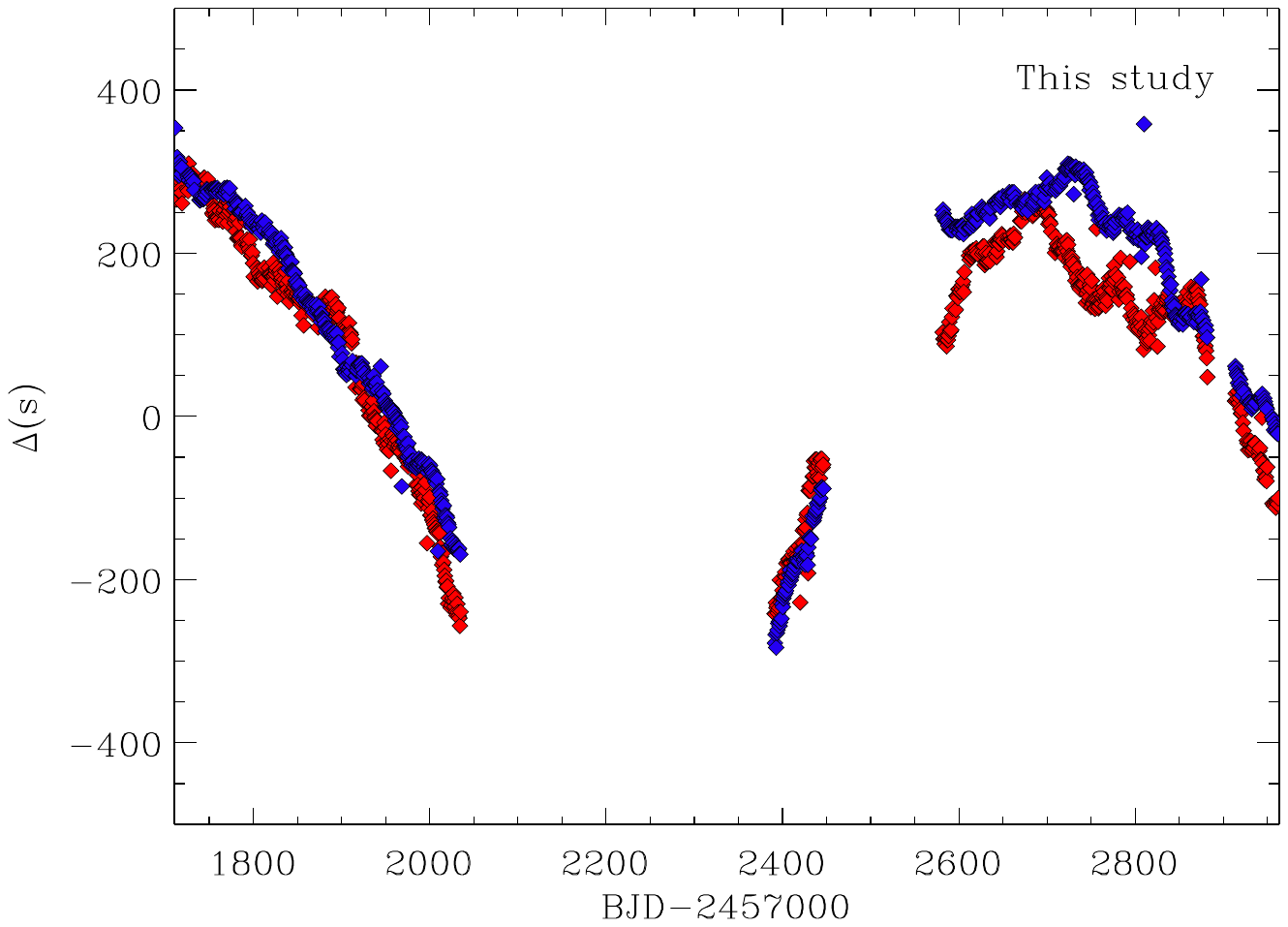}
    \includegraphics[trim = 2.6cm 2.4cm 2.4cm 1.5cm, clip,width=0.48\linewidth]{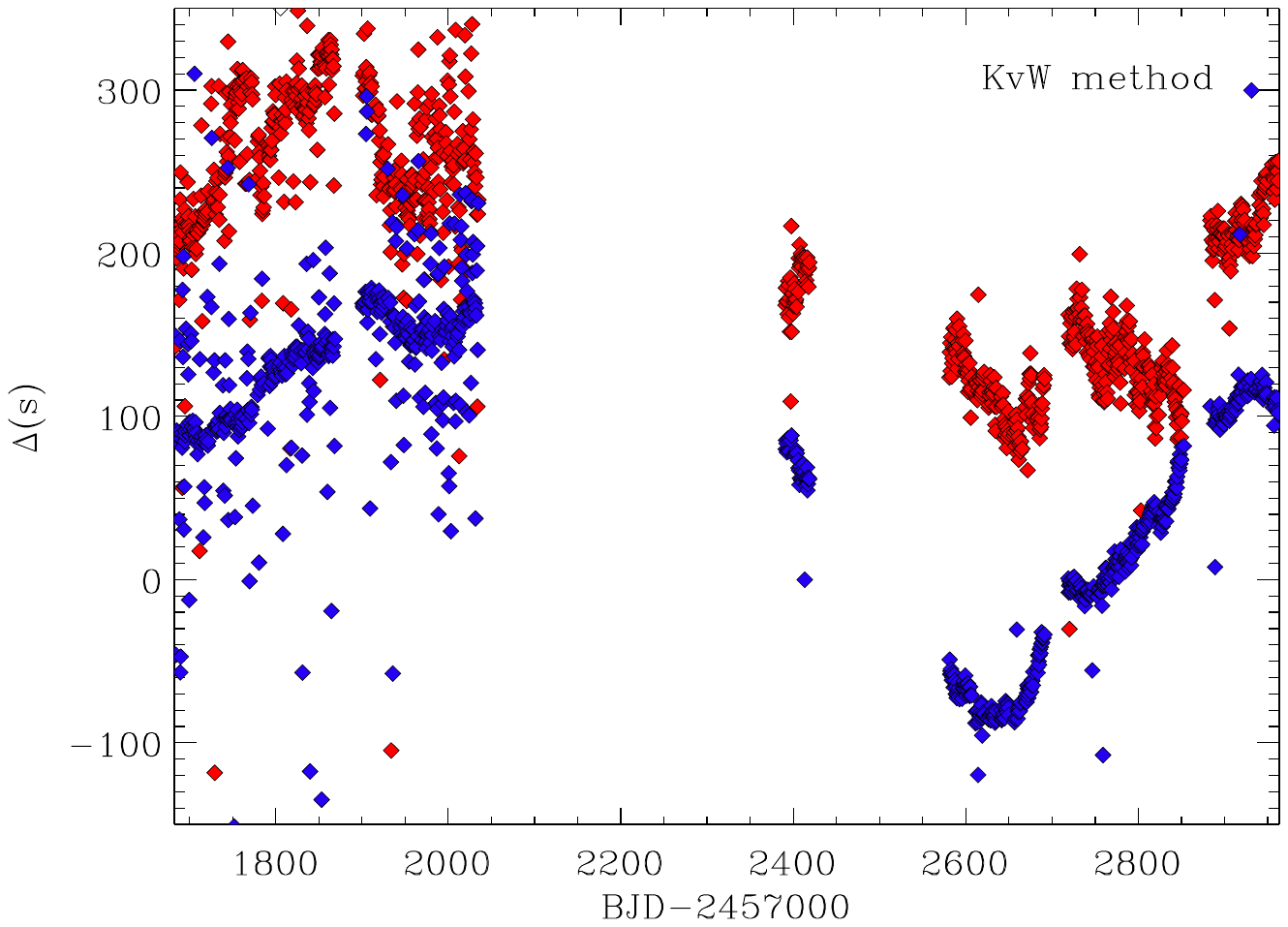}\hspace{0.02\textwidth}%
    \includegraphics[trim = 2.6cm 2.4cm 2.4cm 1.5cm, clip,width=0.48\linewidth]{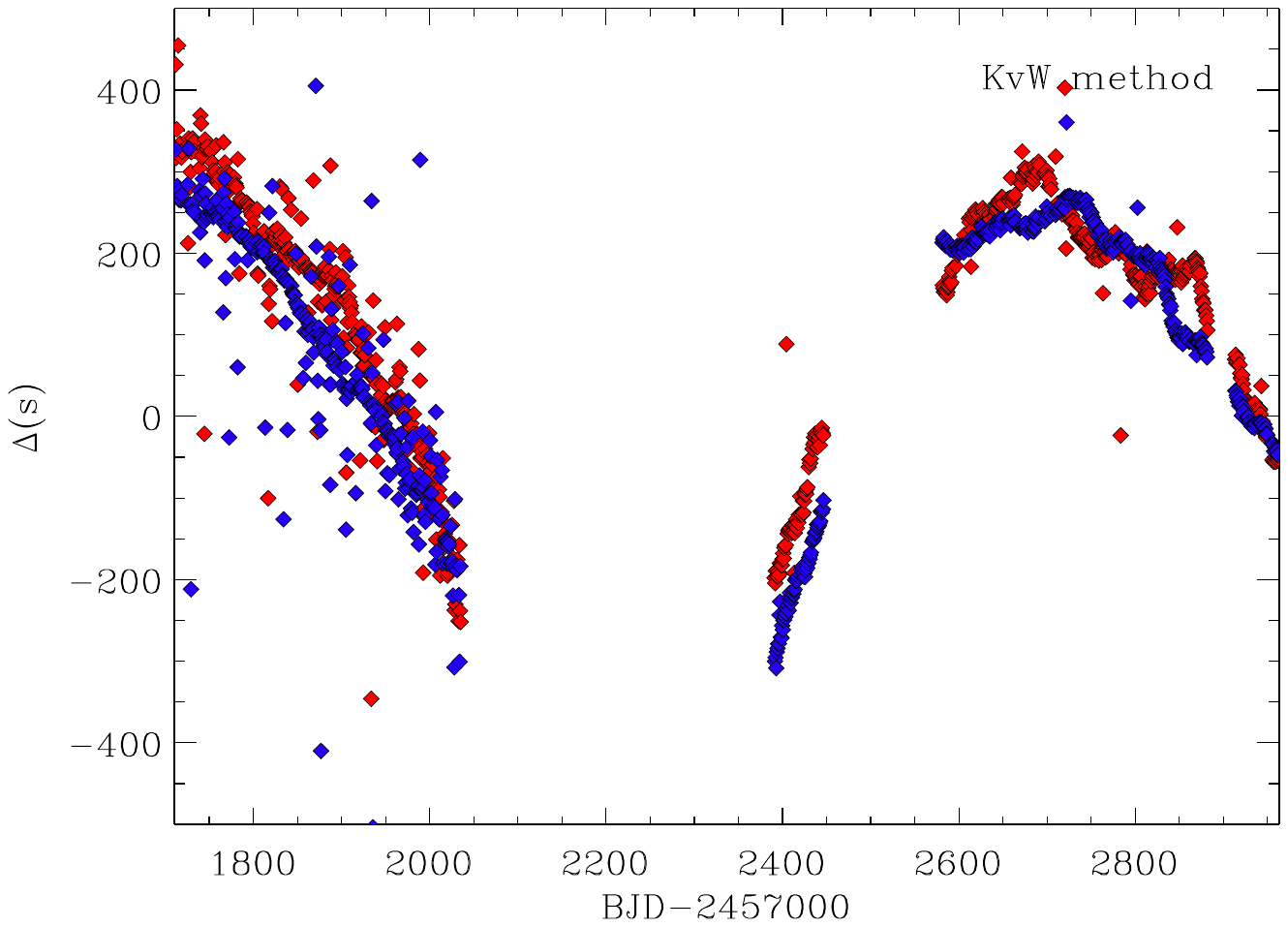}
    \includegraphics[trim = 2.6cm 2.4cm 2.4cm 1.5cm, clip,width=0.48\linewidth]{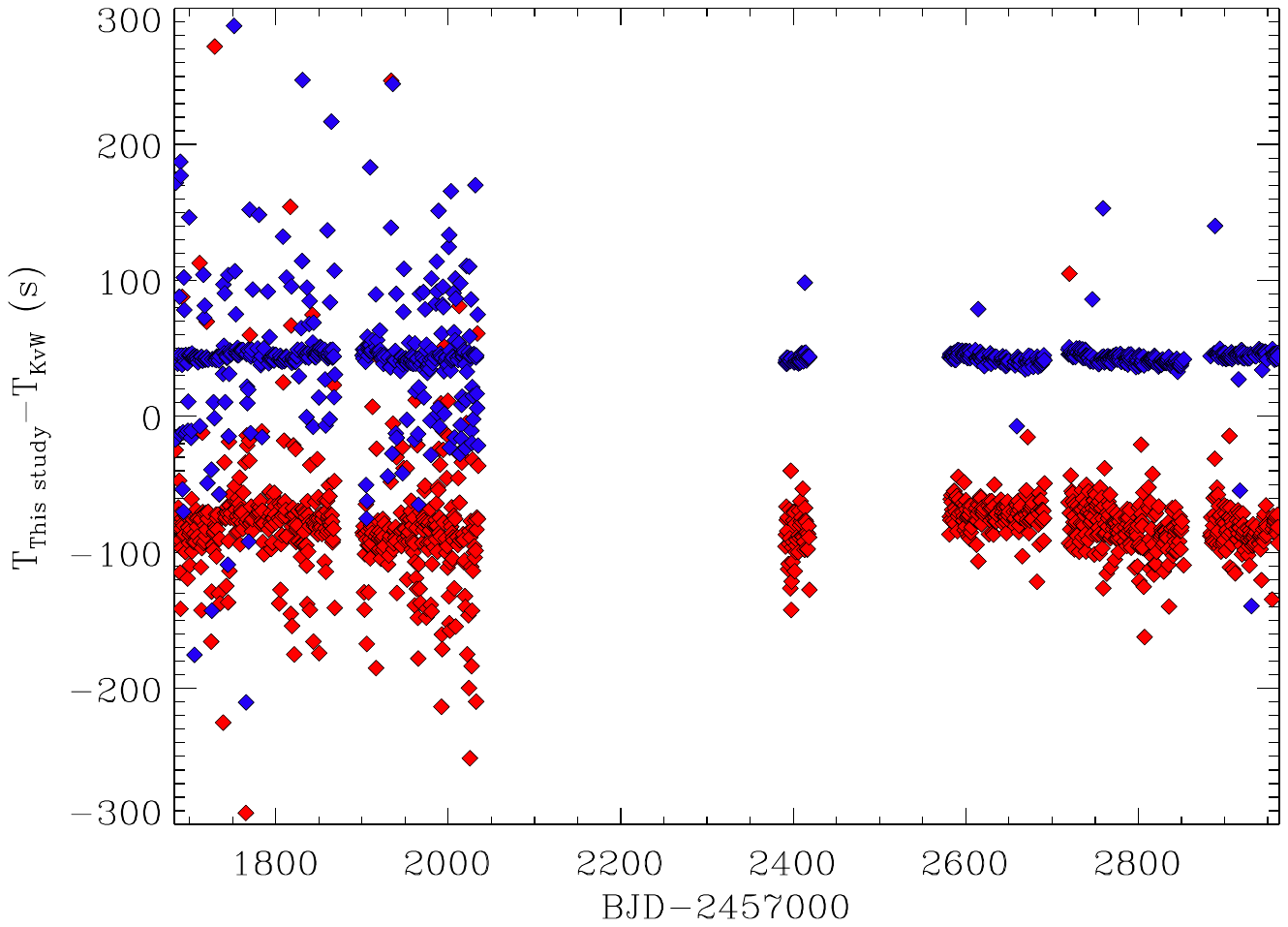}\hspace{0.02\textwidth}%
    \includegraphics[trim = 2.6cm 2.4cm 2.4cm 1.5cm, clip,width=0.48\linewidth]{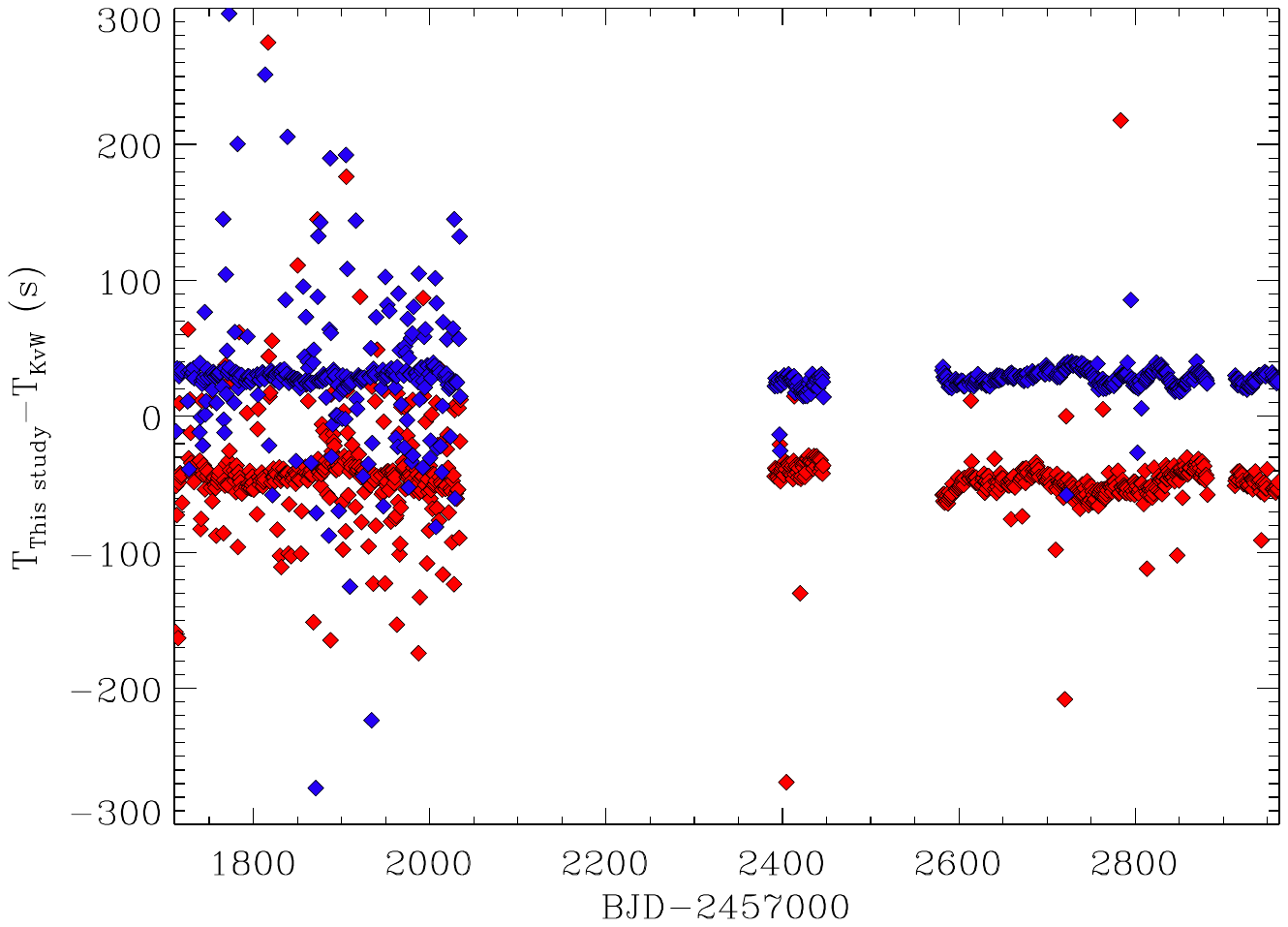}
    \caption{\tbf{Top panels: ETVs for \ticone{} (left) and \ticfive{} (right) extracted using the method presented in this study. Blue and red diamonds refer to the primary and secondary eclipse times, respectively. No outliers have been removed. Middle panels: Same as top panels but using the \kvw{} method. Bottom panels: Difference between the mid-eclipse times measured by the two methods.}}
    \label{fig:kvw}
\end{figure}

\tbf{Due to its simplicity, the classic method of \citeauthor{1956BAN....12..327K}\ (\citeyear{1956BAN....12..327K}; hereafter \kvw{}) has been widely used for the determination of mid-eclipse times since its publication in 1956. This method consists in minimizing the squared sum of the flux differences between data points that are equidistant from an idealized vertical symmetry axis, and assumes that the data points are equally spaced in time and that the eclipse profile is symmetric. In contrast to the \kvw{} method, our template-based method accounts for asymmetry in the eclipse profile of EB light curves that exhibit a significant O'Connell effect. Thus, we decided to investigate the influence of the O'Connell effect on the determination of mid-eclipse times by applying both methods to the SC light curves of \ticone{} and \ticfive{}. For this, we made use of the modified \kvw{} method\footnote{\url{https://github.com/hdeeg/KvW}} of \cite{2020Galax...9....1D}, which provides a more sophisticated treatment of the eclipse timing uncertainties than the original method. To derive the mid-eclipse times, we clipped the light curves and kept only the points that belong to the eclipses, as required by the \kvw{} algorithm. The mid-eclipse times from the two methods are compared in Figure~\ref{fig:kvw}. First, we find that our method produces significantly fewer outliers than the \kvw{} method. We visually inspected the light curves from Sectors 14--26 and noted that they have a larger number of missing points compared to the other sectors, implying that the data from Sectors 14--26 are not perfectly equally spaced. This results in a higher number of outliers obtained with the \kvw{} method before BJD~245$\,$9034 (see middle panels of Figure~\ref{fig:kvw}). Second, we find that the \kvw{} method introduces systematic shifts in the measured mid-eclipse times with respect to those obtained with our method, which are caused by the asymmetry of the light curves \citep{2006Ap&SS.304..363M}. We also note that the primary and secondary times of minima are shifted in opposite directions, resulting in the displacement of the secondary ETV curve with respect to the primary one, as observed in the middle-left panel of Figure~\ref{fig:kvw}. This is consistent with the results obtained by \cite{2024NewA..10902210T} when analyzing the ETV curve of an overcontact system exhibiting the O'Connell effect. Therefore, we caution against using the \kvw{} method for such systems, mainly those that exhibit a rapidly varying O'Connell effect (see \eg \citealt{2023AJ....165..247P}).}

\subsection{\tbf{Comparison with \cite{2024AA...685A..43M}}}

\tbf{Very recently, \citeauthor{2024AA...685A..43M}\ (\citeyear{2024AA...685A..43M}; hereafter \mitnyan{}) reported the detection of 125 new triple candidates from their ETV survey of the \TESS{} northern CVZ, including three of the five systems investigated in this work. This gives us the opportunity to compare our results with those obtained by \mitnyan{} for these three systems, namely \tictwo{} (listed as \tictwoM{} in \mitnyan{}), \ticfour{} (GZ~Dra), and \ticfive{}. The differences are discussed in detail below.}

\begin{enumerate}[label=(\roman*)]
\item \tbf{\ticfour{}: This system was identified by \mitnyan{} as  one of the most strongly inclined triple systems known with \boldm{$i_{\rm m} = 58^\circ \pm 7^\circ$}. The authors measured the mutual inclination by modeling the effect of dynamical perturbations occurring in the system, which are believed to be the cause of the two dips seen in the primary ETV curve around BJD~245$\,$8800 and BJD~245$\,$9755 (see their Figure~3). As expected, we also find these two dips in the primary ETV curve obtained with our timing procedure. However, they are not visible in the secondary ETV curve, which is shown in Figure~\ref{fig:ETV_curve}, at the corresponding epochs. On the other hand, we observe anticorrelated fluctuations between the ETV curves of the primary and secondary eclipses. This is clearly seen in the residuals between BJD \boldm{$\sim$}245$\,$9400 and BJD \boldm{$\sim$}245$\,$9960. We computed the Lomb--Scargle periodograms of the primary and secondary ETV residuals from this portion of the \boldm{$O-C$} diagram and found a period of \boldm{$\sim$160$\,$}d. We point out that the two dips are separated by exactly six times the period value derived above. We conclude that the short-term variations observed for GZ~Dra are more likely due to the presence of spots that migrate in longitude on the stellar surface with a period of about 160$\,$d.}
\item \tbf{\tictwo{} and \ticfive: These two systems were independently identified by \mitnyan{} as CHT candidates with purely LTTE solutions. In order to derive the outer orbital parameters of these systems, they employed a nonlinear Levenberg--Marquardt (LM) optimization algorithm as described in \cite{2015MNRAS.448..946B}. From the derived orbital parameters, they computed the minimum mass of the third component by adopting \boldm{$M_{\rm A} = 2\,{\rm M}_\odot$} and obtained \boldm{$M_{\rm B} = 0.79\,{\rm M}_\odot$} for \tictwo{} and \boldm{$M_{\rm B} = 0.88\,{\rm M}_\odot$} for \ticfive{}. These values are found to be higher by 12 per cent and 40 per cent, respectively, than those reported in this study, which is due to the higher values of \boldm{$a_{\rm A}\sin i_{\rm AB}$} determined by \mitnyan{} from their best-fit models. \mitnyan{} compared the orbital parameters from their ETV analysis with those from the \Gaia{} DR3 non-single-stars (NSS) catalog \citep{2022yCat.1357....0G}. As can be seen in their Figure~12, their ETV solutions tend to have higher values of \boldm{$a_{\rm A}\sin i_{\rm AB}$} than the NSS solutions for systems with incompletely covered outer orbits (see also \citealt{2023A&A...670A..75C}), which is consistent with the discrepancies observed between their results and ours. We also note that their ETV solutions have higher outer eccentricities than ours, namely \boldm{$e_{\rm AB} = 0.50 \pm 0.01$} for \tictwo{} and \boldm{$e_{\rm AB} = 0.38 \pm 0.02$} for \ticfive{}, implying a possible correlation between the outer eccentricity and the projected semi-major axis. In the context of ETV studies, it was shown by \cite{2013ApJ...768...33R} that the LM method is not suitable for exploring parameter space with strong nonlinear covariances between parameters. This may explain the discrepancies with respect to the results of \mitnyan{}.}
\end{enumerate}%

\subsection{Comparison between SC and FFI results}\label{sec:comparison}

\begin{figure}
    \centering
    \includegraphics[trim = 3.2cm 2.7cm 4.2cm 2.0cm,clip,width=0.55\columnwidth,angle=0]{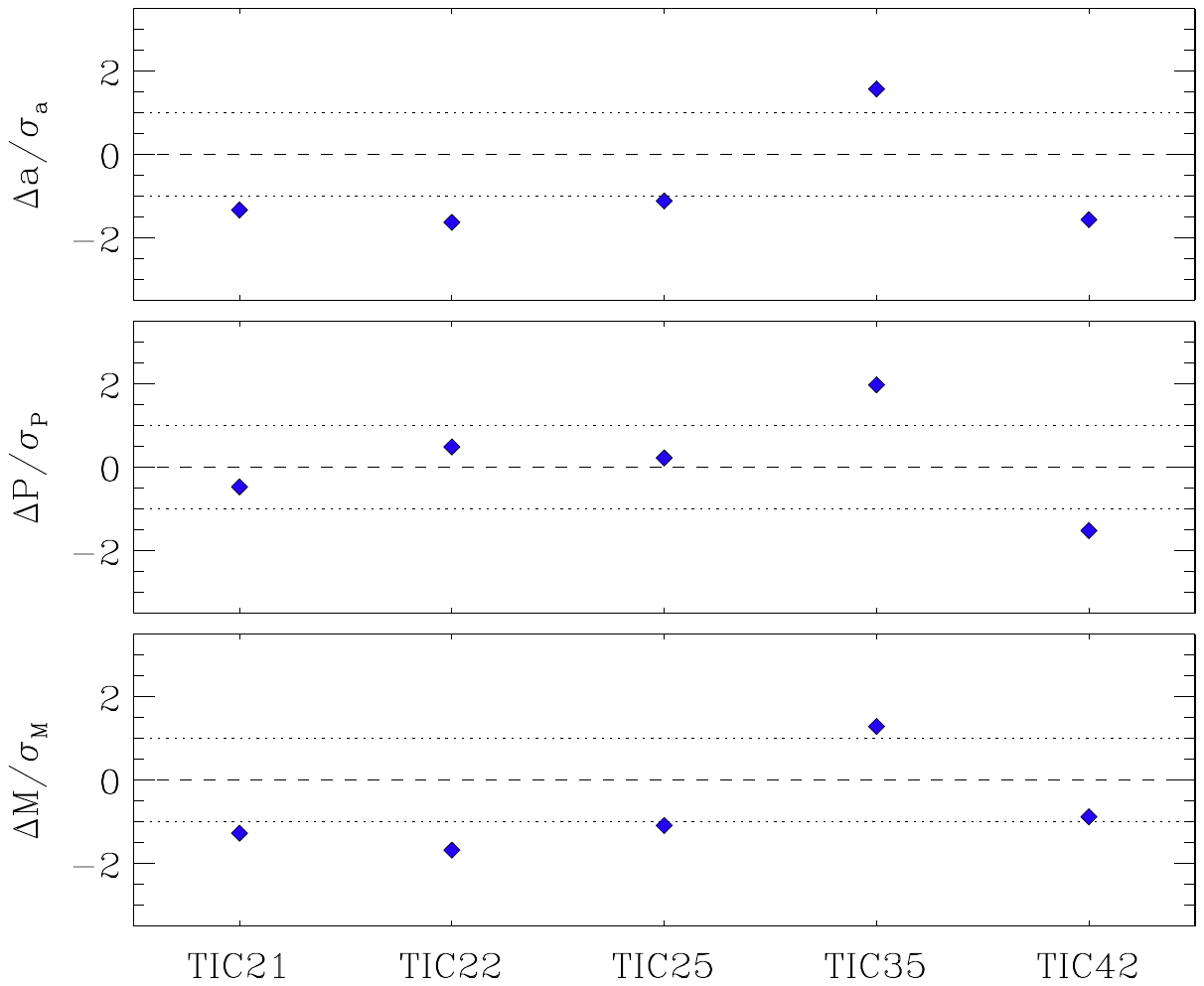}
    \caption{Difference in standard deviations between the system parameter estimates obtained from the two datasets. The top, middle, and bottom panels show the difference in $a_{\rm A}\sin i_{\rm AB}$, $P_{\rm AB}$, and $M_{\rm B}$, respectively.}
    \label{fig:param_diff}
\end{figure}

In Section~\ref{sec:etv}, we summarized the ETV analysis of five newly detected triple systems using the SC and FFI datasets independently. This allowed us to derive the orbital parameters of each system as listed in Table~\ref{tab:ETV_param} and Appendix~\ref{sec:ETV_param}. In principle, the mass of the third body can be obtained from Kepler's third law applied to the EB's barycentric orbit, namely:
\begin{equation}
\frac{M_{\rm B}^3}{(M_{\rm A}+M_{\rm B})^2} = \frac{a_{\rm A}^3}{P_{\rm AB}^2},
\end{equation}
where $M_{\rm A} = M_{\rm Aa} + M_{\rm Ab}$ is the sum of the masses of the eclipsing components Aa and Ab. Here, the projected semi-major axis $a_{\rm A}\sin i_{\rm AB}$ is related to $A_{\rm LTTE}$ by Equation~(\ref{eq:Altte}). We can then write the mass function of the third body as
\begin{equation}
f(M_{\rm B}) = \frac{M_{\rm B}^3 \, \sin^3 i_{\rm AB}}{(M_{\rm Aa}+M_{\rm Ab}+M_{\rm B})^2} = 1.074 \times 10^{-3} \, \frac{A_{\rm LTTE}^3}{P_{\rm AB}^2},\label{eq:fm}
\end{equation}
where the masses are expressed in the units of solar mass, $A_{\rm LTTE}$ is in seconds, and $P_{\rm AB}$ is in days. Assuming an inclination of $i_{\rm AB}\!=90^\circ$, Equation~(\ref{eq:fm}) provides a lower limit on the mass $M_{\rm B}$ of the third component. Unfortunately, the masses of the eclipsing components, $M_{\rm Aa}$ and $M_{\rm Ab}$, are not presently known for the systems studied in this paper. Thus, for our calculations, we randomly drew $900\,000$ values of $M_{\rm Aa}$ and  $M_{\rm Ab}$, corresponding to 90 per cent of the chain length after rejection of the burn-in phase, from a normal distribution centered on 1.00$\,{\rm M}_\odot$ with a standard deviation of $0.01\,{\rm M}_\odot$. We solved the mass function for the minimum tertiary mass by means of Laguerre's method \citep{1992nrca.book.....P}, using the posterior distribution samples of $A_{\rm LTTE}$ and $P_{\rm AB}$ obtained from our MCMC fitting in Section~\ref{sec:etv}. We then computed the median and the 16 per cent and 84 per cent credible intervals on the minimum tertiary mass from the corresponding posterior probability distribution. For each system, we listed in Table~\ref{tab:masses} the values of $a_{\rm A}\sin i_{\rm AB}$, $P_{\rm AB}$, and $M_{\rm B}$ derived from both the SC and FFI datasets, and we compared them in Figure~\ref{fig:param_diff}.\\

\begin{table}
    \centering
    \begin{minipage}{135mm}
        \caption{Derived values of $a_{\rm A}\sin i_{\rm AB}$, $P_{\rm AB}$, and $M_{\rm B}$ for the five systems.}
        \label{tab:masses}
        {
        \renewcommand{\arraystretch}{1.0}
        \begin{tabular}{@{}lcccc@{}}
            \hline
            \hline
            TIC & $a_{\rm A}\sin i_{\rm AB}$ & $P_{\rm AB}$ & $M_{\rm B} \, (i_{\rm AB}\!=90^\circ$) & Obs.\ mode \\
                & (au)                       & (d)          & (M$_\odot$)                            &            \\
            \hline
            \multirow{ 2}{*}{219900027} & $\mbf{0.2280^{+0.0014}_{-0.0014}}$ & $\mbf{1159.4^{+3.3}_{-2.1}}$   & $\mbf{0.1773^{+0.0015}_{-0.0014}}$ & SC  \\
                                        & $\mbf{0.208^{+0.015}_{-0.017}}$    & $\mbf{1144.1^{+32.4}_{-39.0}}$ & $\mbf{0.162^{+0.012}_{-0.013}}$    & FFI \\
            \hline
            \multirow{ 2}{*}{229771234} & $0.360^{+0.011}_{-0.010}$    & $359.8^{+0.8}_{-0.8}$    & $0.706^{+0.026}_{-0.025}$    & SC  \\
                                        & $0.316^{+0.027}_{-0.026}$    & $361.7^{+4.0}_{-3.9}$    & $0.600^{+0.063}_{-0.056}$    & FFI \\
            \hline
            \multirow{ 2}{*}{259006185} & $0.0773^{+0.0027}_{-0.0027}$ & $199.6^{+0.5}_{-0.5}$    & $0.1953^{+0.0072}_{-0.0072}$ & SC  \\
                                        & $0.0716^{+0.0051}_{-0.0048}$ & $199.8^{+0.9}_{-0.9}$    & $0.180^{+0.014}_{-0.013}$    & FFI \\
            \hline
            \multirow{ 2}{*}{356896561} & $0.1026^{+0.0022}_{-0.0022}$ & $477.5^{+2.1}_{-2.0}$    & $0.1425^{+0.0033}_{-0.0033}$ & SC  \\
                                        & $0.1084^{+0.0037}_{-0.0037}$ & $485.2^{+4.5}_{-3.9}$    & $0.1493^{+0.0055}_{-0.0053}$ & FFI \\
            \hline
            \multirow{ 2}{*}{424461577} & $\mbf{0.6430^{+0.0031}_{-0.0032}}$ & $\mbf{995.6^{+3.4}_{-2.9}}$    & $\mbf{0.6274^{+0.0045}_{-0.0046}}$ & SC  \\
                                        & $\mbf{0.607^{+0.023}_{-0.019}}$    & $\mbf{965.5^{+19.8}_{-16.3}}$  & $\mbf{0.600^{+0.031}_{-0.027}}$    & FFI \\
            \hline 
        \end{tabular}
        }
    \end{minipage}
    \tablecomments{The mass of the third component, $M_{\rm B}$, was computed assuming $i_{\rm AB}\!=90^\circ$ and $M_{\rm Aa} = M_{\rm Ab} = 1.00 \pm 0.01 \, {\rm M}_\odot$ (see the text for details).}
\end{table}

Based on our analysis of the SC and FFI data, we derived the minimum tertiary mass for each system with a precision better than 4 per cent and \tbf{11} per cent, respectively. Furthermore, as can be seen in Figure~\ref{fig:param_diff}, the minimum mass value determined from the FFI data is underestimated (by more than $\sim$\tbf{0.9}$\sigma$) for four of the five systems considered. The minimum mass is related to the semi-major axis and thus to $A_{\rm LTTE}$ via Kepler's third law, implying that the latter was underestimated during the fitting procedure (see top panel of Figure~\ref{fig:param_diff}).
According to the morphological classification described in \cite{2012AJ....143..123M}, our target sample can be divided into three categories: detached (\ticthree{} and \ticfour{}), semi-detached (\ticone{} and \ticfive{}), and contact (\tictwo{}). For the two detached binaries, \ticthree{} and \ticfour{}, we were able to reach a precision on the minimum tertiary mass better than 8 per cent using the FFI data, with a mass discrepancy lower than 1.3$\sigma$. The rms of the primary and secondary ETV residuals are of the order of 20 and 40$\,$s, respectively. We obtained comparable results for the semi-detached binary \ticfive{}, despite the large scatter of the ETVs (rms of $\sim$200$\,$s). The ETV amplitude observed for this system ($A_{\rm LTTE} \simeq$ \tbf{316}$\,$s) is indeed high enough to allow us to properly constrain the minimum tertiary mass. The two remaining binaries, \ticone{} and \tictwo{}, have the highest morphology values in our sample, \ie 0.684 and 0.722, respectively. Their ETV curves are affected by a large scatter of \tbf{140}--210$\,$s for the primary and \tbf{300}--\tbf{320}$\,$s for the secondary. As a consequence, both the precision ($\gtrsim\,$\tbf{8} per cent) and accuracy ($\gtrsim\,$\tbf{1.3}$\sigma$) of the minimum tertiary mass are degraded. For the three latter binaries, we found that the change of the FFI cadence from 30 to 10 minutes improves the quality of the timing significantly. This can be seen in Figures~\ref{fig:ETV_tic21}, \ref{fig:ETV_tic22}, and \ref{fig:ETV_tic42} in Appendix~\ref{sec:ETV_curve} by comparing the scatter of the ETVs before and after BJD~245$\,$9060, which corresponds to the last day of the 30-min cadence observations (Sector 27). It is thus expected that the new 200-s FFI cadence will significantly improve our ability to detect and characterize stellar and substellar companions around the shortest-period EBs. Although based on a small number of objects, these results highlight the importance of using the calibrated SC data to validate the ETV analysis of FFI targets. 

\subsection{Global properties of the systems}

In this section, we discuss the global properties of each of the five investigated systems based on our results and the literature.

\subsubsection{TIC 259006185}

\ticthree{} is a detached EB consisting of two unequal mass components in a $\sim$1.9-d circular orbit. Its light curve exhibits shallow secondary eclipses that were discarded from the final analysis due to their low signal-to-noise ratio. The LTTE solution suggests the presence of a minimum 0.20$\,{\rm M}_\odot$ companion orbiting the eclipsing pair with a short period of about 200$\,$d. In such a compact orbital configuration, the EB system is expected to experience significant dynamical perturbations from the third body. Using Equations~(\ref{eq:Adyn}) and~(\ref{eq:Altte}), we computed the ratio of the amplitudes of the dynamical and LTTE effects, ${\mathcal A}_{\rm dyn}/{\mathcal A}_{\rm LTTE}$, which is found to be 0.6. Such a result implies that the dynamical contribution is no longer negligible with respect to the LTTE contribution. While our LTTE model is sufficient to compare the ETV curves from the two datasets, a more detailed analysis is required to properly constrain the minimum mass of the third body. We also point out that the observed ETV curve of \ticthree{} can be explained by the presence of migrating spots on the surface of either or both stars. Indeed, it was shown by \cite{2013ApJ...774...81T} that migrating spots can produce ETVs with typical periods of the order of 50--200$\,$d and induce an anticorrelated behavior of the primary and secondary minima variations. Most of our systems exhibit such spot-induced ETVs, as can be seen in the residuals of Figures~\ref{fig:ETV_curve}, \ref{fig:ETV_tic21}, \ref{fig:ETV_tic22}, and \ref{fig:ETV_tic42}. Unfortunately, in the absence of precise ETV measurements of the secondary eclipses, it is impossible to determine the origin of the primary ETV signal observed for \ticthree{}.

\subsubsection{TIC 229771234 and TIC 424461577}

As a result of our analysis, we identified two new compact hierarchical triples with outer periods less than 1000$\,$d, namely \tictwo{} ($P_{\rm AB}=359.8\,$d) and \ticfive{} ($P_{\rm AB}=\mbf{995.6}\,$d). For both systems, the minimum mass of the third component (0.6--0.7$\,{\rm M}_\odot$) is consistent with that of a main-sequence K-type star. As noted by \cite{2014AJ....147...87T}, CHTs are rare among the entire population of hierarchical triples. They are thought to form differently than wider triple systems, most likely through a sequential disk fragmentation mechanism \citep{2020MNRAS.491.5158T,2021Univ....7..352T}. From their ETV analysis of \Kepler{} targets, \cite{2016MNRAS.455.4136B} provided the largest sample of short-outer-period triples to date, which contains 104 CHTs. In particular, they noted a deficit of short-period EBs ($P_{\rm A} \lesssim 1\,$d) with a close tertiary companion ($P_{\rm AB} \lesssim 200\,$d), \tbf{recently confirmed by \cite{2024AA...685A..43M}}, suggesting additional differences between the formation of the tightest binaries and that of their longer-period counterparts. In order to better constrain the formation mechanisms of close binaries, it is thus important to increase the sample of well-characterized CHTs, especially those with binary periods $\lesssim\,$1$\,$d such as \tictwo{} ($P_{\rm A}=0.821\,$d) and \ticfive{} ($P_{\rm A}=0.745\,$d). The example of \tictwo{} demonstrates the potential of our method for recovering ETV signals with semi-amplitudes as low as 180$\,$s (typical rms of 210$\,$s) using the FFIs. For the shortest-period EBs, the ETV amplitude induced by a stellar companion in a 50-d orbit is expected to be $\sim$50$\,$s \citep{2016MNRAS.455.4136B}, i.e., of the same order as the scatter of the 
ETV measurements obtained for \tictwo{} and \ticfive{} from the SC data (rms of 20--90$\,$s). Thus, detecting the tightest CHTs (those with $P_{\rm A} \lesssim 1\,$d and $P_{\rm AB} \lesssim 200\,$d) represents a real challenge that requires a systematic follow-up, based on \TESS{} SC photometry, of the most promising candidates.\\ 

For completeness, we also computed the expected radial velocity (RV) semi-amplitude induced by the companion on the EB, which is defined as\\
\begin{equation}
K_{\rm A} = \frac{21.80 \, A_{\rm LTTE}}{P_{\rm AB} \, (1-e_{\rm AB}^2)^{1/2}},
\end{equation}
where $K_{\rm A}$ is in units of km\,s$^{-1}$, $A_{\rm LTTE}$ is in units of seconds, and $P_{\rm AB}$ is in units of days. We found the RV semi-amplitudes for \tictwo{} and \ticfive{} to be easily measurable with modern spectrographs, namely $K_{\rm A} = 11.6$ and 7.1$\,$km\,s$^{-1}$, respectively. By fitting a double-Keplerian orbit, it will then be possible to constrain the semi-amplitude of the ETV signal, the individual masses of the eclipsing components and, thus, the minimum mass of the circumbinary companion (see \eg \citealt{2020MNRAS.499.3019M}). \tbf{Additionally, the tertiary mass can be directly measured by performing a combined analysis of the RV and interferometric measurements of the outer relative orbit (see \eg \citealt{2018A&A...617A...2M}). Using Kepler's third law, we estimated the semi-major axis of the outer relative orbit to be 19.4~mas for \tictwo{} and 6.4~mas for \ticfive{}. We conclude that these two systems can be spatially resolved with long-baseline interferometry, which has proven its ability to resolve close binary stars with separations of a few mas \citep{2016A&A...586A..35G,2019A&A...632A..31G,2023A&A...672A.119G}.} 

\subsubsection{TIC 219900027}

\ticone{} (AU~Dra) is known as a semi-detached EB with an orbital period of 0.515$\,$d \citep{1998IBVS.4587....1B,1998IBVS.4605....1S}. 
We identified the companion as being a low-mass red dwarf, with a minimum mass of about \tbf{0.18}$\,$M$_\odot$, orbiting the eclipsing pair in a $\sim$3.2-yr eccentric orbit ($e_{\rm AB} =$ \boldm{$0.5055^{+0.0029}_{-0.0055}$}). Additionally, a total of 69 times of minima were recorded in the $O-C$ gateway\footnote{\url{http://var2.astro.cz/ocgate/}} \citep{2006OEJV...23...13P} for the period 1956--2017, thereby allowing us to examine the long-term behavior of the measured ETVs. From the archival times of minima, we found that the orbital period of the eclipsing pair is continuously decreasing, which is suggested by a negative parabolic trend in the corresponding $O-C$ diagram. To account for the observed parabolic trend, a quadratic term, $c_2 E^2$, has to be included in Equation~(\ref{eq:OC_mod}). The rate of period change can then be estimated from the relation $\dot{P} = 2c_2/P$, where $c_2$ is a fitted parameter. Orbital period changes are commonly observed in close EBs \citep{2022AJ....163..157H}, and may result either from the effects of a magnetic activity cycle, known as the Applegate mechanism \citep{1992ApJ...385..621A}, or from the mass transfer between the binary components. The long-term period decrease observed for \ticone{} may indicate that the EB is undergoing mass transfer accompanied by angular momentum loss, leading to a shrinkage of its orbit. A detailed period-change analysis of the system, which is beyond the scope of this paper, could shed some light on its past and future evolution.

\subsubsection{TIC 356896561}

Among the five studied systems, \ticfour{} has the tertiary companion with the lowest minimum mass, namely $M_{\rm B} = 0.143\,{\rm M}_\odot (= 149\,{\rm M}_{\rm Jup})$. For our calculations, we assumed the individual masses of the eclipsing components to be $M_{\rm Aa} = M_{\rm Ab} = 1.00 \pm 0.01 \, {\rm M}_\odot$. If we consider now a low-mass EB with component masses of $M_{\rm Aa} = M_{\rm Ab} = 0.300 \pm 0.003 \, {\rm M}_\odot$, we find that the minimum tertiary mass is $M_{\rm B} = 68.5 \pm 1.6 \, {\rm M}_{\rm Jup}$, which is below the hydrogen-burning mass limit of $\sim$80$\,{\rm M}_{\rm Jup}$ that separates brown dwarfs (BDs) from very low-mass stars \citep{2002A&A...382..563B}. \ticfour{} may therefore consist of a low-mass EB orbited by a possible BD-mass companion in a $\sim$1.3-yr orbit. This demonstrates the feasibility of detecting circumbinary companions with masses as low as those of BDs around low-mass EBs using the FFIs. Furthermore, the EB orbit is expected to undergo dynamical perturbations caused by the close-orbiting low-mass companion, as previously mentioned in Section~\ref{sec:etv}. \cite{2016MNRAS.455.4136B} suggested a threshold value of ${\mathcal A}_{\rm dyn}/{\mathcal A}_{\rm LTTE} \sim 0.25$ above which a combined dynamical and LTTE model may be required. We derived the expected amplitude of the dynamical ETV contribution from Equation~(\ref{eq:Adyn}) by adopting $M_{\rm A} = 0.6\,{\rm M}_\odot$ and $M_{\rm B} = 0.065\,{\rm M}_\odot (= 68.5\,{\rm M}_{\rm Jup})$. We obtained ${\mathcal A}_{\rm dyn} \simeq 14\,$s, leading to a ratio ${\mathcal A}_{\rm dyn}/{\mathcal A}_{\rm LTTE} \sim 0.27$, which is very close to the threshold value. In addition, the dynamical amplitude is found to be of the same order as the rms scatter in the ETV residuals ($\sim$10--20$\,$s for the SC dataset), implying that the dynamical contribution to the ETVs cannot be properly assessed. Therefore, we limited our analysis to a simple LTTE model.\\

\ticfour{} (WDS~18127+5446) is classified as a visual binary in the Washington Double Star Catalog (WDS\footnote{\url{https://crf.usno.navy.mil/wds}}; \citealt{2001AJ....122.3466M}). This suggests that the EB+BD system may be gravitationally bound to a fourth body that corresponds to the secondary component of the visual pair. We attempted to determine the orbital period of the fourth body by using the position measurements found in the literature. We approximated the  angular semi-major axis of the visual pair as the mean of the measured angular separations listed in Table~\ref{tab:rel_pos}. We then computed the absolute semi-major axis by adopting the trigonometric parallax of $4.84 \pm 1.66\,$mas from \cite{1997AA...323L..49P}, and found it to be $\sim$78$\,$au. Finally, adopting this value and assuming a total mass of 3$\,{\rm M}_\odot$, we derived an orbital period of about 400$\,$years for the visual pair from Kepler's third law, making confirmation of its gravitationally-bound nature impossible. We also found eight additional eclipse times in the $O-C$ gateway and saw no evidence of a long-term trend caused by a fourth body.

\begin{table}
    \centering
    \begin{minipage}{110mm}
        \caption{Relative positions of the two visual components of \ticfour{}.}
        \label{tab:rel_pos}
        {
        \renewcommand{\arraystretch}{1.0}
        \begin{tabular}{@{}cccc@{}}
            \hline
            \hline
            Time        & Angle      & Separation & Reference                  \\
            Bess.\ yr.\ & ($^\circ$) & (\arcsec)  &                            \\
            \hline
            1983.42     & 355.1      & 0.370      & \cite{1984AAS...57..467M}  \\
            1983.52     & 348.6      & 0.380      & \cite{1984AAS...57..467M}  \\
            1983.54     & 343.7      & 0.380      & \cite{1984AAS...57..467M}  \\
            1985.50     & 348.7      & 0.330      & \cite{1987ApJS...65..161H} \\
            1991.25     & 343.0      & 0.428      & \cite{1997AA...323L..49P}  \\
            1991.73     & 346.8      & 0.450      & \cite{2002AA...384..180F}  \\
            1992.10     & 341.6      & 0.340      & \cite{1997AAS..126..273M}  \\
            1995.64     & 347.9      & 0.350      & \cite{1996ApJS..105..475H} \\
            \hline 
        \end{tabular}
        }
    \end{minipage}
\end{table}

\section{Summary and future prospects}\label{sec:summary}

In this study, we aimed to explore the feasibility of performing precise eclipse timing of \TESS{} EBs using the FFIs. To this end, we developed a fast, automated method, based on a Bayesian approach, and applied it successfully to a sample of $\sim$100 EBs selected from the Villanova \TESS{} EB catalog. After visual inspection of the $O-C$ diagrams, we identified a total of ten triple system candidates with outer periods less than $\sim$1300$\,$d. For five of them, we were able to properly constrain the outer orbit from the SC and FFI data independently, thanks to the good coverage of the photometric observations in each mode, and to derive the minimum mass of the third component with a precision better than 4 per cent and \tbf{11} per cent, respectively. This approach allowed us to compare the results obtained from the two datasets. Thus, we found that using the FFI data results in the degradation of both the accuracy and precision of the tertiary mass determination for the tightest EBs and in the underestimation of the third component's mass. However, we stress that the main results obtained from the FFI data regarding the nature of the detected signals are consistent with those obtained from the SC data. Our main findings and prospects are as follows.
\begin{enumerate}
    \item CHTs are rare, and their detection is challenging as the amplitude of the ETV signal decreases with decreasing outer period. This was pointed out by \cite{2016MNRAS.455.4136B}, who noticed the almost complete absence of short-period EBs with a close tertiary companion (\ie with $P_{\rm AB} \lesssim 200\,$d) among their sample of 222 \Kepler{} triple candidates. Furthermore, due to their compactness, CHTs are of particular interest for our understanding of the formation \citep{2021Univ....7..352T} and evolution \citep{2020A&A...640A..16T,2022A&A...661A..61T} of hierarchical triple stellar systems. In this study, we report the detection of two new CHTs, \tictwo{} and \ticfive{}, with respective outer periods of 359.8 and \tbf{995.6}$\,$d. For \tictwo{}, we were able to recover the low-amplitude signal of the close companion from the FFIs, despite the large scatter of the ETV measurements. Based on these results, we may expect to  find more of these systems by conducting a systematic search for ETVs among \TESS{} FFI EBs. Additionally, we argue that detecting the tightest CHTs will require a dedicated photometric survey based on \TESS{} SC data. In this context, the \Gaia{} 
    DR3 
    NSS catalog 
    represents a powerful tool for identifying the most promising CHT candidates, as recently shown by \cite{2023A&A...670A..75C}.
    
    \item The BD desert refers to the low occurence of BD companions around main-sequence stars with periods shorter than $\sim$5$\,$yr \citep{2006ApJ...640.1051G}. This desert is particularly pronounced in the mass range 35$-$55$\,{\rm M}_{\rm Jup}$ and might indicate different formation mechanisms for close-orbiting BDs and isolated ones \citep{2014MNRAS.439.2781M}. BDs are generally believed to form like stars, through the fragmentation of molecular clouds \citep{2004ApJ...617..559P}. However, \cite{2014MNRAS.439.2781M} suggested the existence of two  populations of BD companions, distinguishable by their physical and orbital properties, as recently confirmed by \cite{2023MNRAS.526.5155S} from a larger sample of BDs. One of them consists of high-mass BDs formed via molecular cloud fragmentation, while the other consists of low-mass BDs that are thought to be formed by disk instability in the same manner as planets \citep{1997Sci...276.1836B}. In this work, we identified a possible BD companion orbiting the eclipsing binary \ticfour{} with a relatively short period of $\sim$1.3$\,$yr, placing it in the BD desert. Based on their spectroscopic survey of 47 single-lined EBs, \cite{2019A&A...624A..68M} reported the absence of circumbinary BDs with orbital periods less than 6000$\,$d, suggesting the existence of a BD desert around binary stars that needs to be confirmed. Using numerical simulations, \cite{2021MNRAS.504.4291G} investigated the potential of the ETV method in detecting BD-mass companions in close orbits and concluded that the BD desert is not due to detection limits. Therefore, to improve our understanding of the BD desert, we aim to increase the number of confirmed circumbinary BDs by taking advantage of the large sample of EBs expected to be detected in the \TESS{} FFIs.
    
    \item In addition to the LTTE, there are several other effects that can produce ETVs, such as starspot migration, apsidal motion, and mass transfer. Spot-induced ETVs have typical periods of $\sim$50--200$\,$d and show a distinctive anticorrelated pattern between the primary and secondary minima variations, as indicated by \cite{2013ApJ...774...81T}. Most of our EBs exhibit such spot-induced variations, and we suspect that the $\sim$200-d signal found in the $O-C$ diagram of \ticthree{} may be caused by the presence of migrating starspots rather than by a third body. \cite{2015MNRAS.448..429B} presented a detailed analysis of 414 \Kepler{} binaries with anticorrelated primary and secondary ETVs. They investigated the effect of differential stellar rotation on the relative motions of spots and found that the differential rotation of stars in tight binaries is less pronounced than that of single stars. The reason for this difference remains to be understood. In contrast to the spot-induced variations, ETVs caused by apsidal motion or mass transfer occur on long time-scales and require decades of observations from ground- and space-based telescopes to be detected. The analysis of these long-term variations has important implications for a variety of studies. For example, by measuring the apsidal motion in eccentric EBs, it is possible to test general relativity \citep{2021A&A...649A..64B} and stellar structure models \citep{2021A&A...654A..17C}. In the framework of the Contact Binaries Towards Merging (CoBiToM) Project, \cite{2021MNRAS.502.2879G} and \cite{2022MNRAS.514.5528L} investigated the evolution of ultra-short-period contact binaries through the analysis of their $O-C$ curves, which were obtained during a decades-long monitoring program. These examples highlight the importance of extending the time span of the observations using the high-precision photometric data provided by the \TESS{} mission.\\
\end{enumerate}%


\noindent%
This paper includes data collected by the \TESS{} mission, which are publicly available from the MAST. Funding for the \TESS{} mission is provided by NASA’s Science Mission directorate. This research has made use of NASA’s Astrophysics Data System Bibliographic Services, the SIMBAD data base, operated at CDS, Strasbourg, France and the VizieR catalogue access tool, CDS, Strasbourg, France. The original description of the VizieR service was published in \cite{2000A&AS..143...23O}.

We gratefully acknowledge support from the NASA TESS Guest Investigator grant 80NSSC22K0180 (PI: A.~Pr\v{s}a).

\tbf{Finally, we also thank the anonymous referee for comments that helped us to improve this paper.}

%

\vspace{5mm}
\facilities{\TESS{}.
    }


\software{\astroquery{} \citep{2019AJ....157...98G},
    \eleanor{} \citep{2019PASP..131i4502F}.
    }



\clearpage
\appendix

\section{Light curves}\label{sec:LC}

\begin{figure}[h!]
    \centering
    \includegraphics[trim = 2.9cm 4.0cm 2.4cm 2.8cm,clip,width=0.5\columnwidth,angle=0]{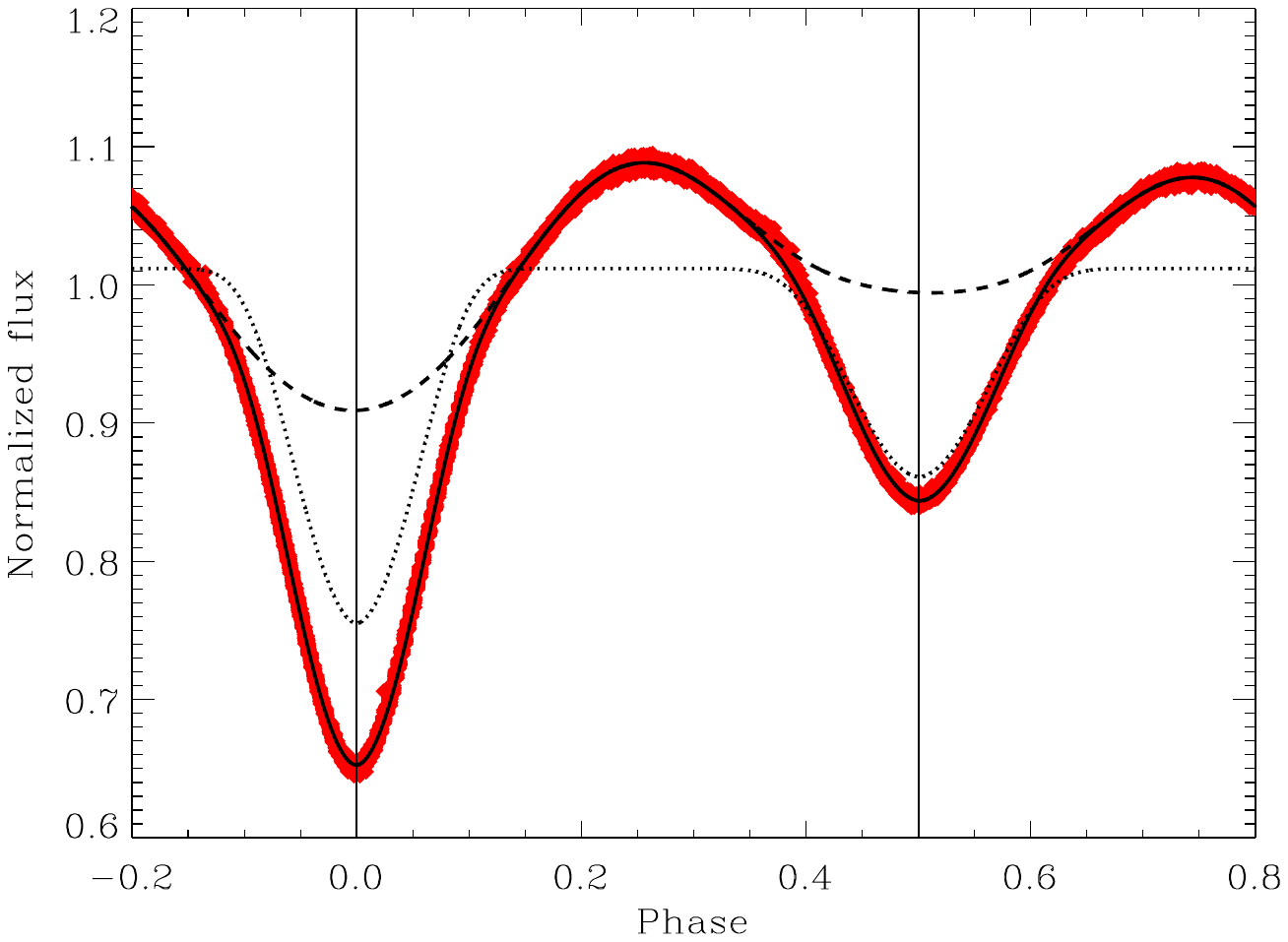}
    \includegraphics[trim = 2.9cm 2.4cm 2.4cm 2.8cm,clip,width=0.5\columnwidth,angle=0]{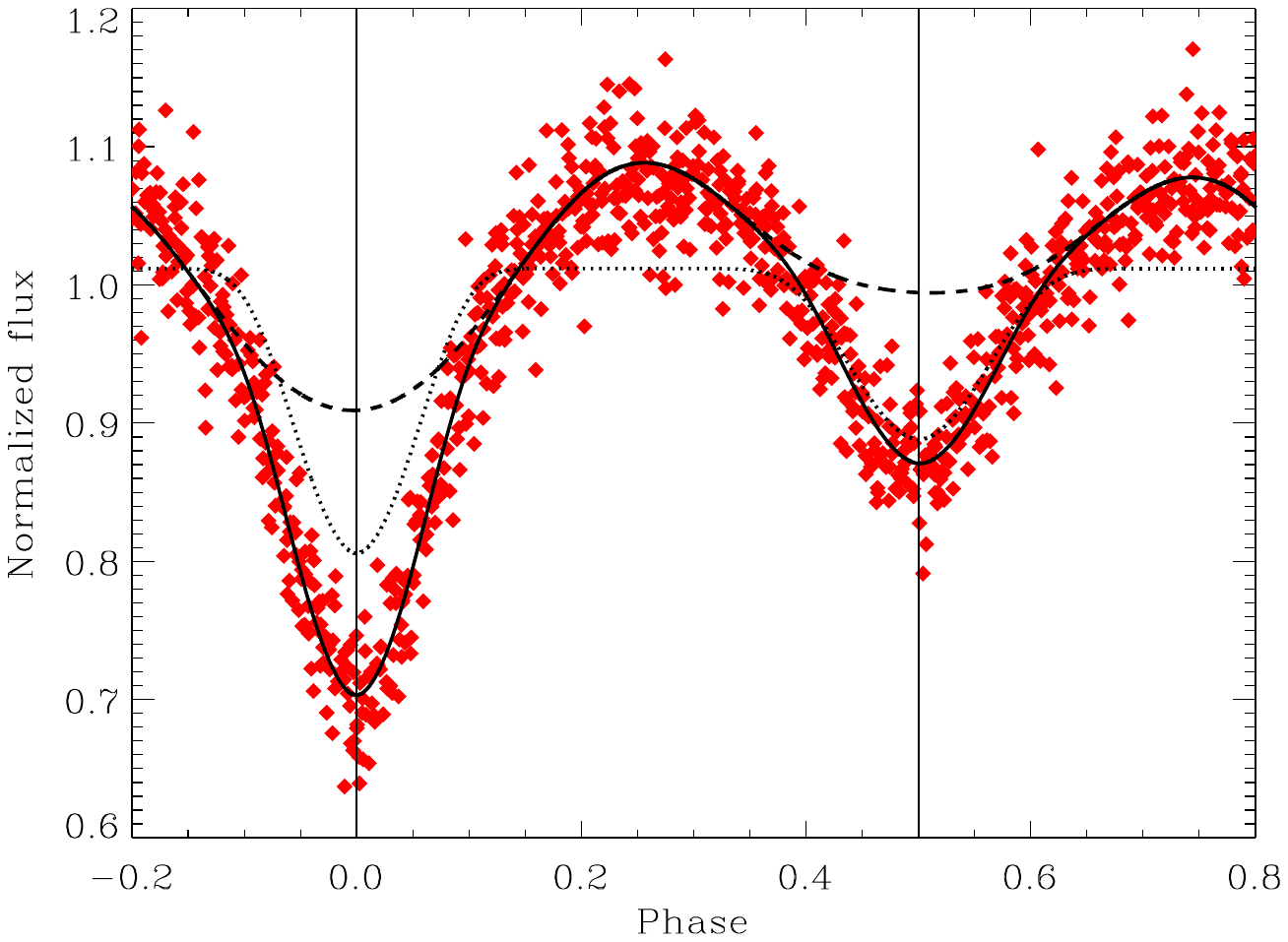}
    \caption{Same as Figure~\ref{fig:LC} but for \ticone{} (presented here is Sector~19). \tbf{The eclipse profile and the contribution of the O'Connell and proximity effects are shown as the dotted and dashed lines, respectively. To account for the eclipse depth difference between the SC and FFI light curves, we derived an independent light-curve solution for each of the two datasets.}}
\end{figure}

\begin{figure}[h!]
    \centering
    \includegraphics[trim = 2.9cm 4.0cm 2.4cm 2.8cm,clip,width=0.5\columnwidth,angle=0]{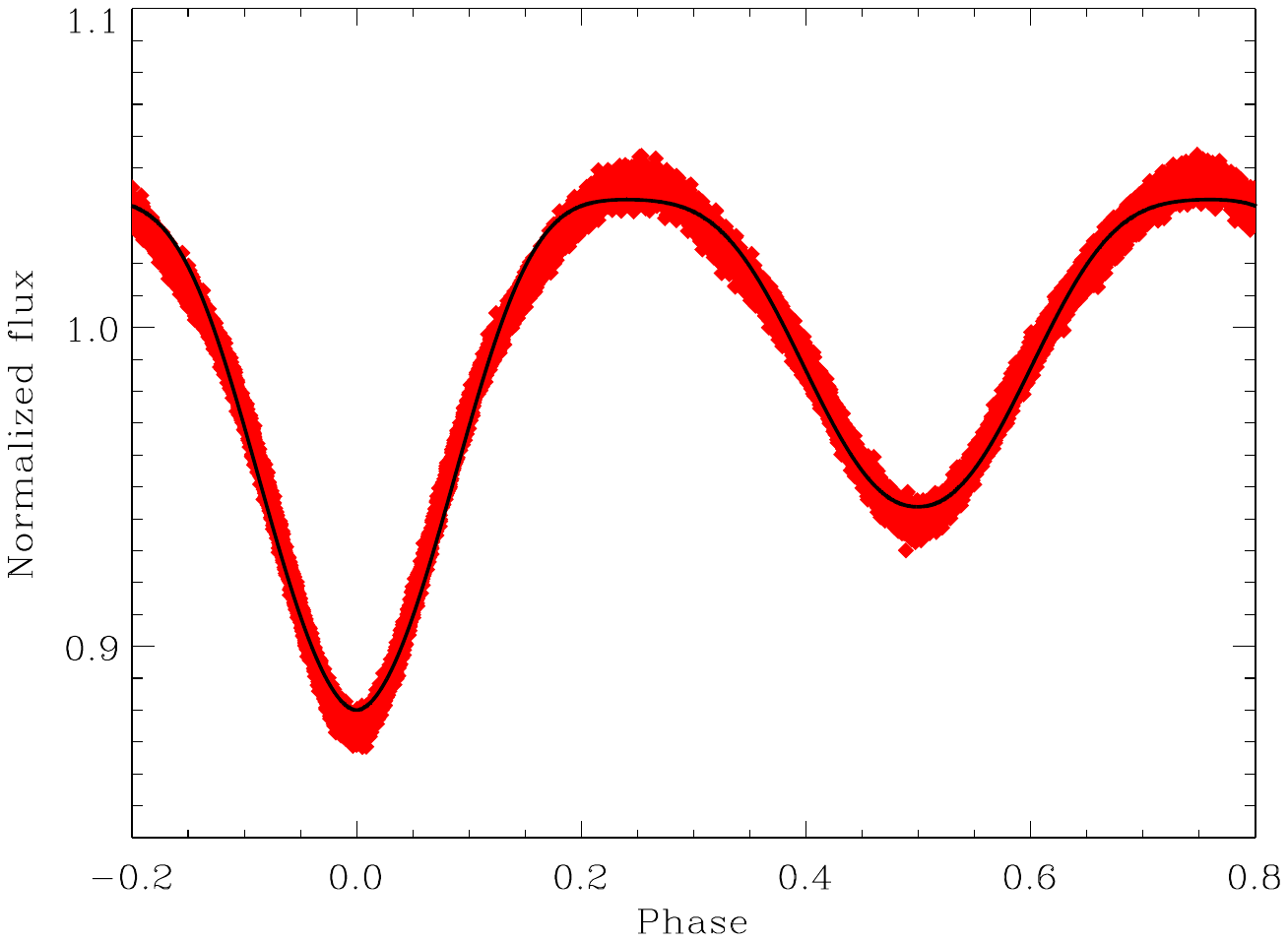}
    \includegraphics[trim = 2.9cm 2.4cm 2.4cm 2.8cm,clip,width=0.5\columnwidth,angle=0]{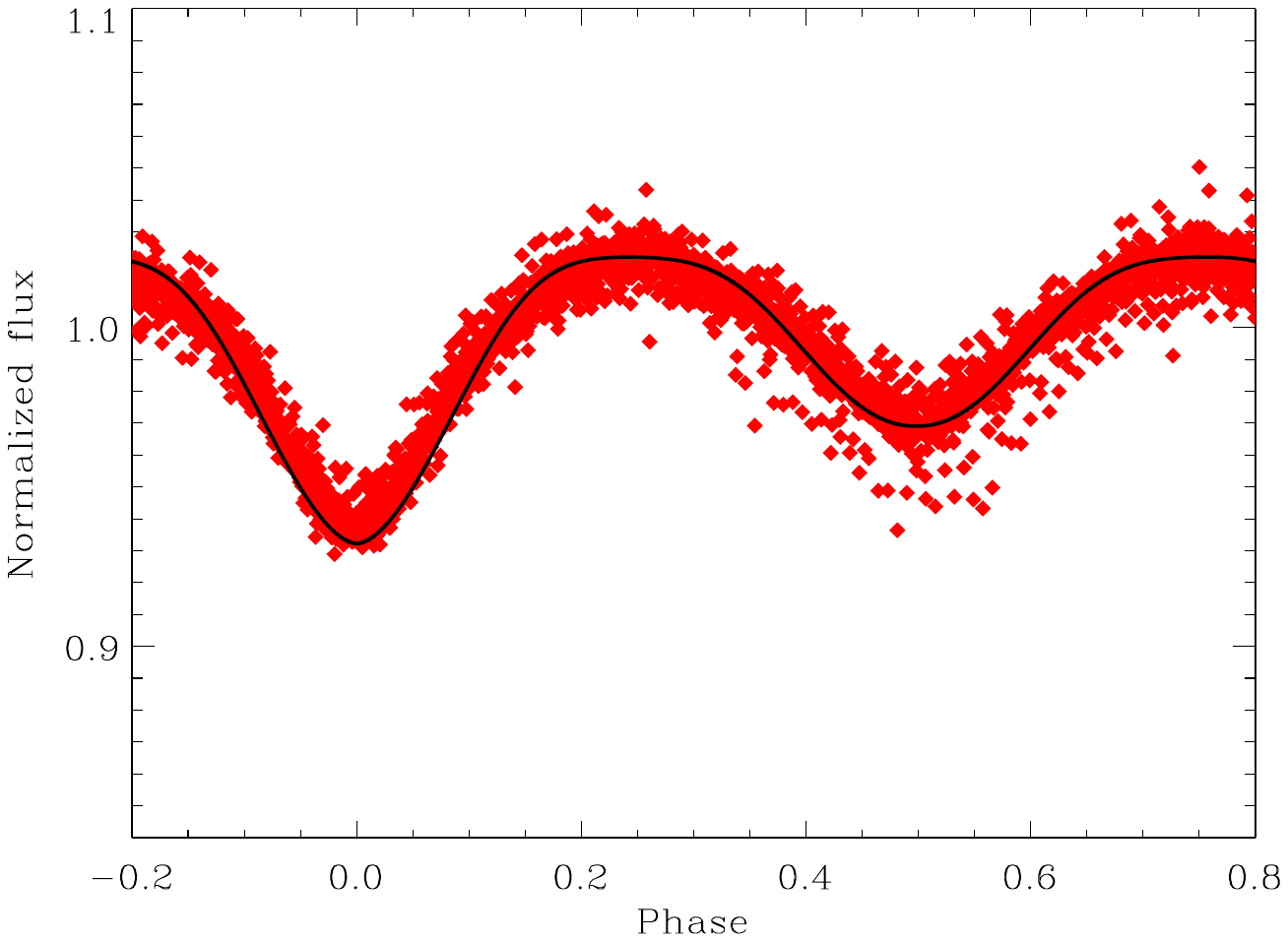}
    \caption{Same as Figure~\ref{fig:LC} but for \tictwo{} (presented here is Sector~52). To account for the eclipse depth difference between the SC and FFI light curves, we derived an independent light-curve solution for each of the two datasets.}
\end{figure}

\begin{figure}[h!]
    \centering
    \includegraphics[trim = 2.9cm 4.0cm 2.4cm 2.8cm,clip,width=0.5\columnwidth,angle=0]{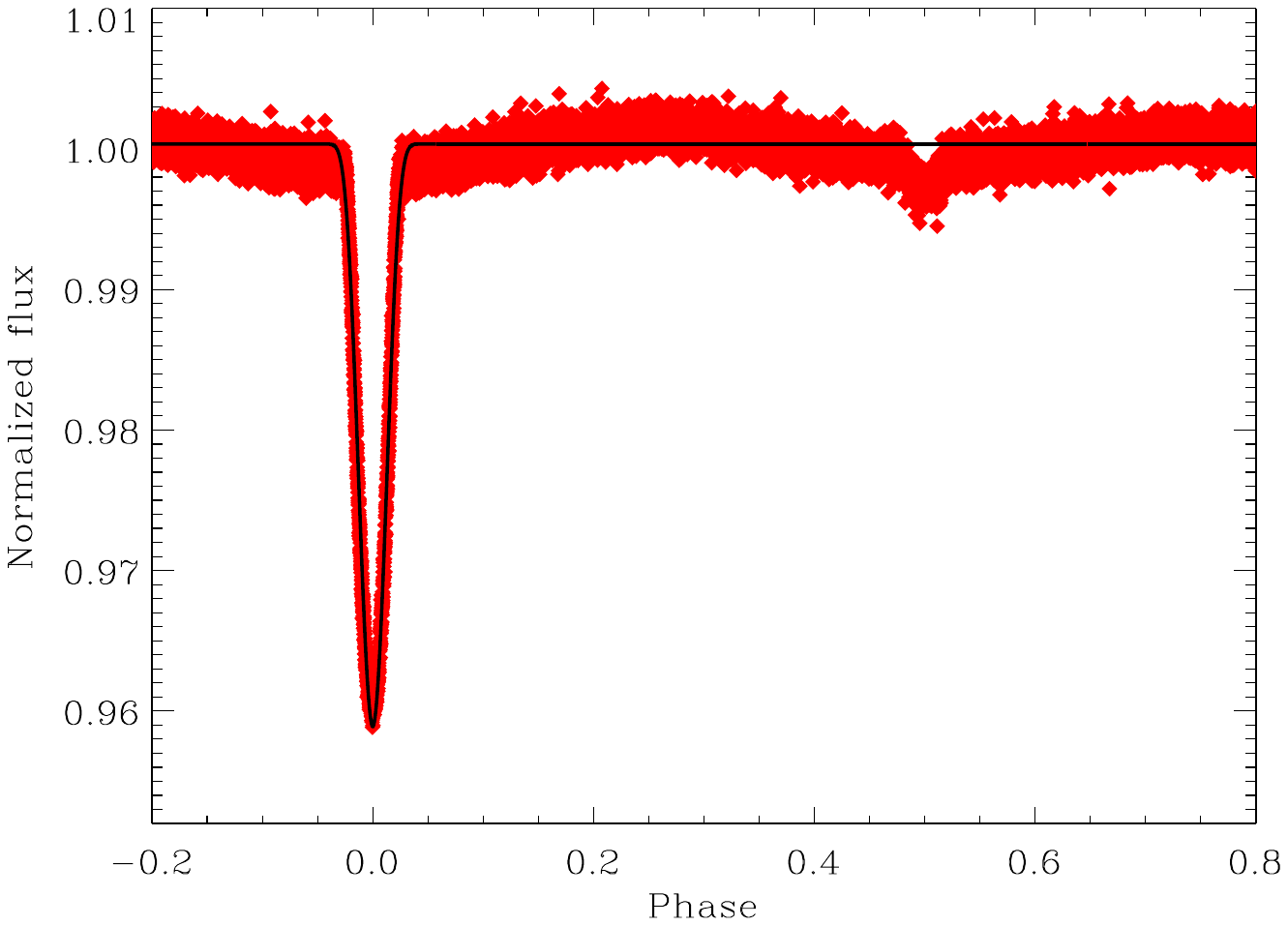}
    \includegraphics[trim = 2.9cm 2.4cm 2.4cm 2.8cm,clip,width=0.5\columnwidth,angle=0]{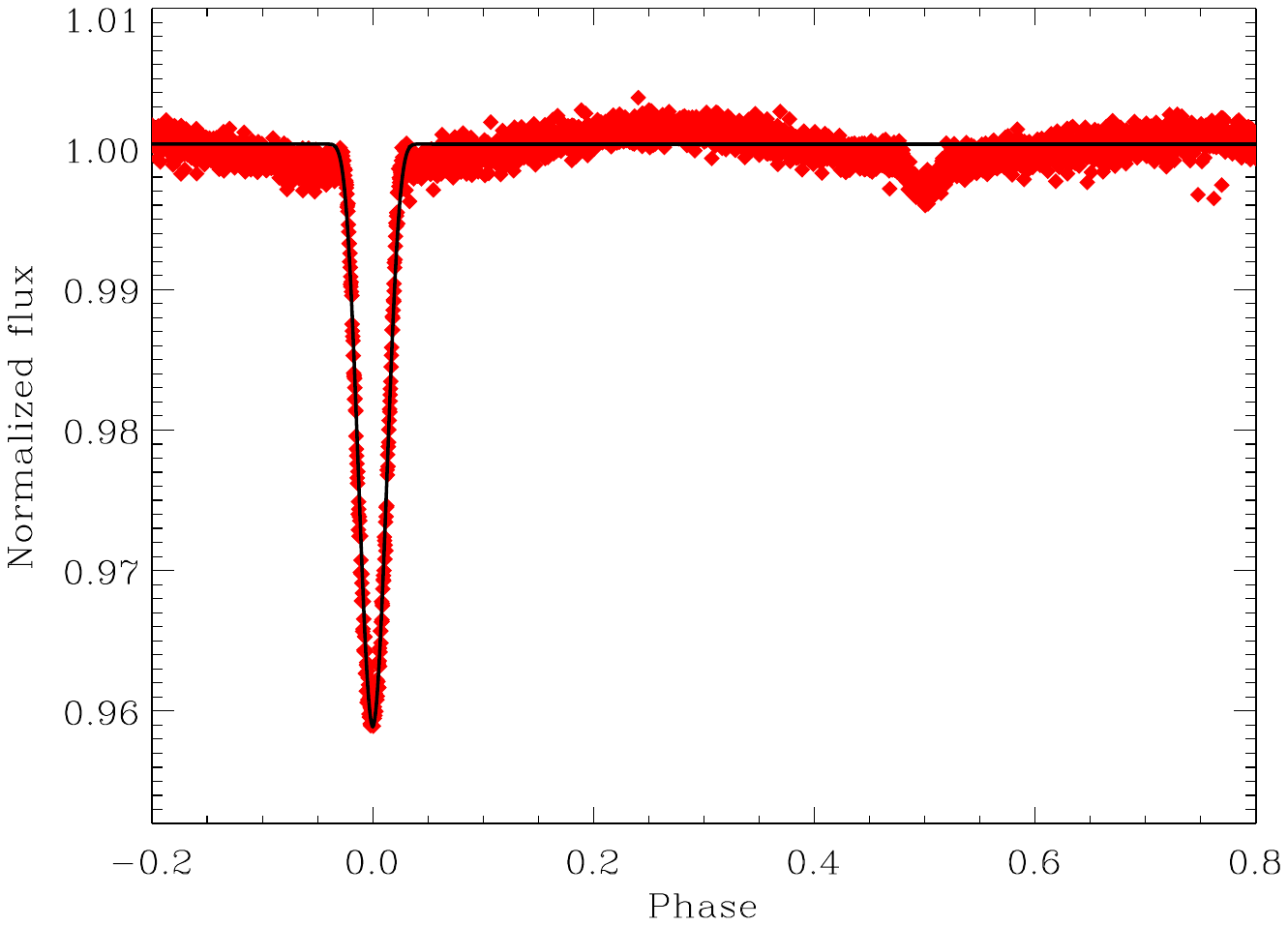}
    \caption{Same as Figure~\ref{fig:LC} but for \ticthree{} (presented here is Sector~55).}
\end{figure}

\begin{figure}[h!]
    \centering
    \includegraphics[trim = 2.9cm 4.0cm 2.4cm 2.8cm,clip,width=0.5\columnwidth,angle=0]{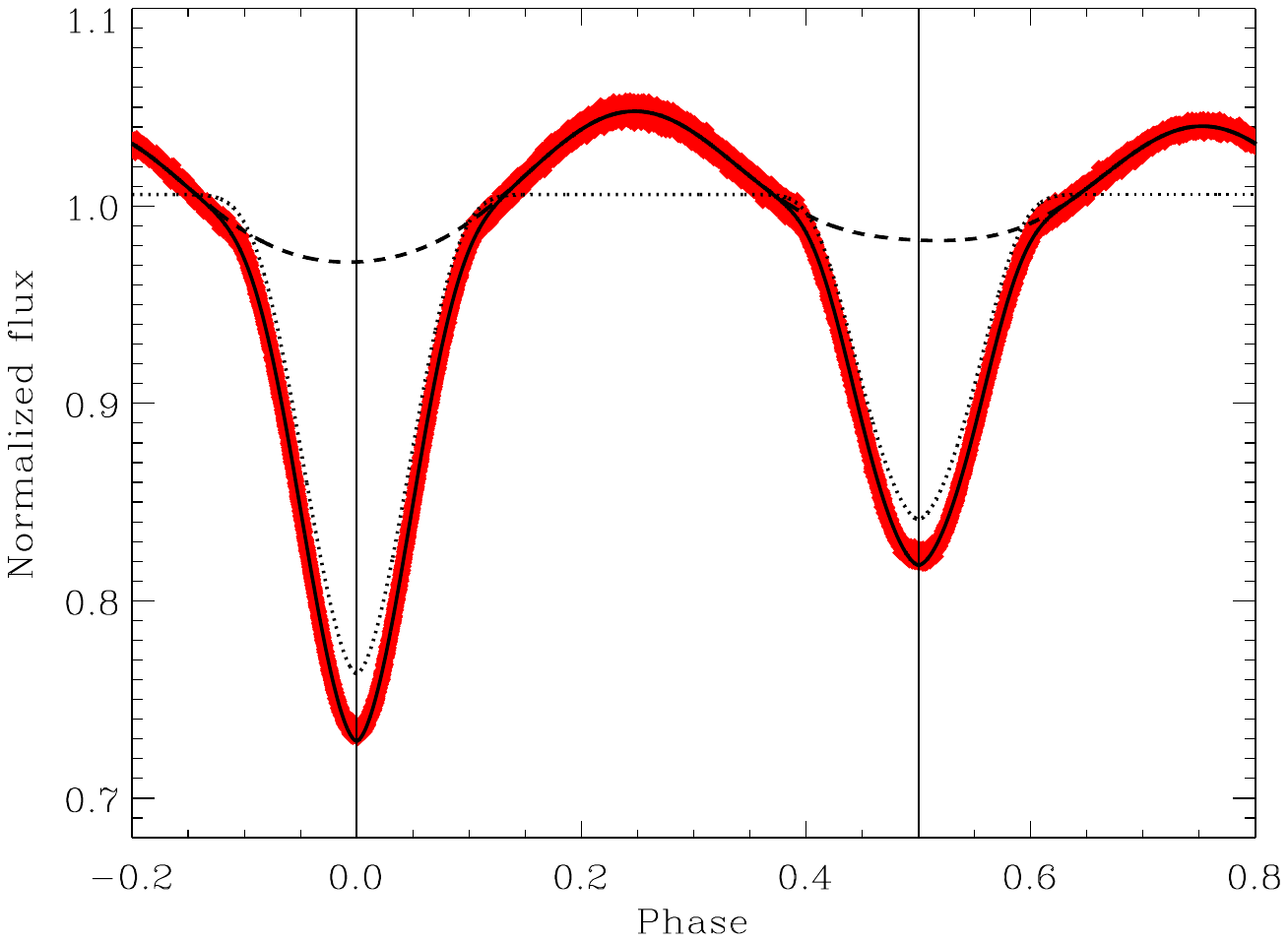}
    \includegraphics[trim = 2.9cm 2.4cm 2.4cm 2.8cm,clip,width=0.5\columnwidth,angle=0]{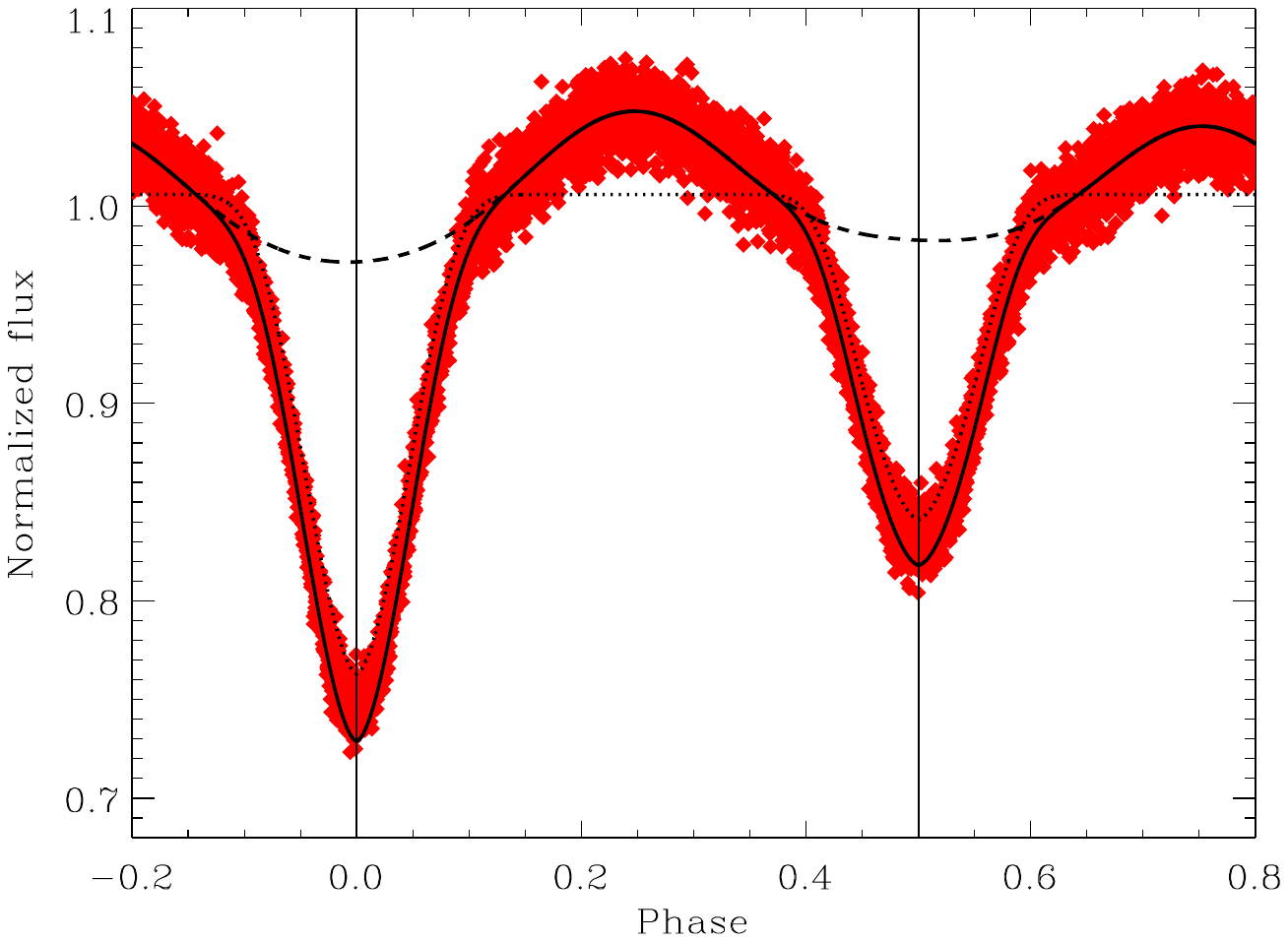}
    \caption{Same as Figure~\ref{fig:LC} but for \ticfive{} (presented here is Sector~57). \tbf{The eclipse profile and the contribution of the O'Connell and proximity effects are shown as the dotted and dashed lines, respectively.}}
\end{figure}

\clearpage
\section{Eclipse timing variation diagrams}\label{sec:ETV_curve}

\begin{figure}[h!]
    \centering
    \includegraphics[trim = 2.0cm 3.7cm 3.5cm 1.5cm,clip,width=0.6\columnwidth,angle=0]{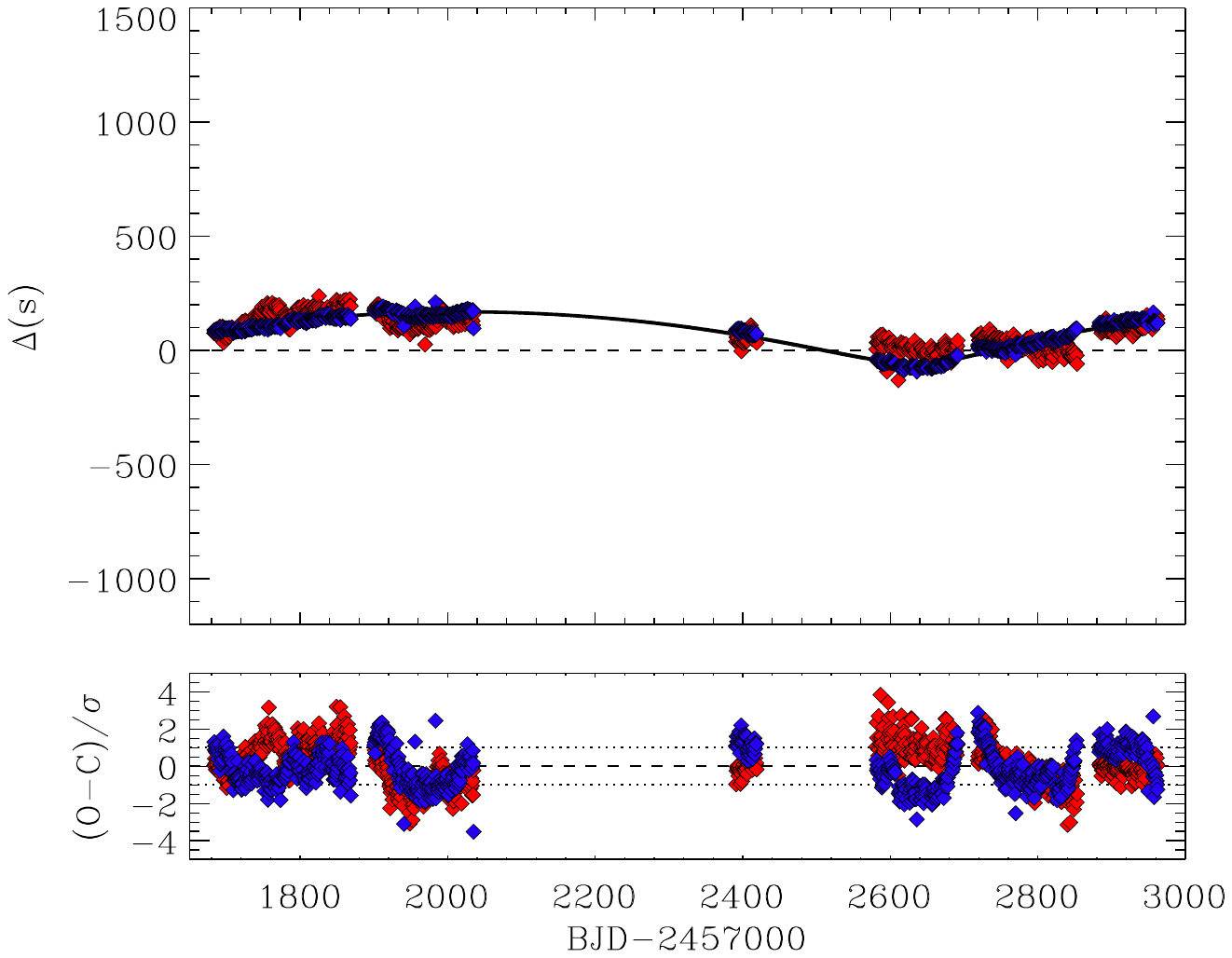}
    \includegraphics[trim = 2.0cm 2.0cm 3.5cm 1.5cm,clip,width=0.6\columnwidth,angle=0]{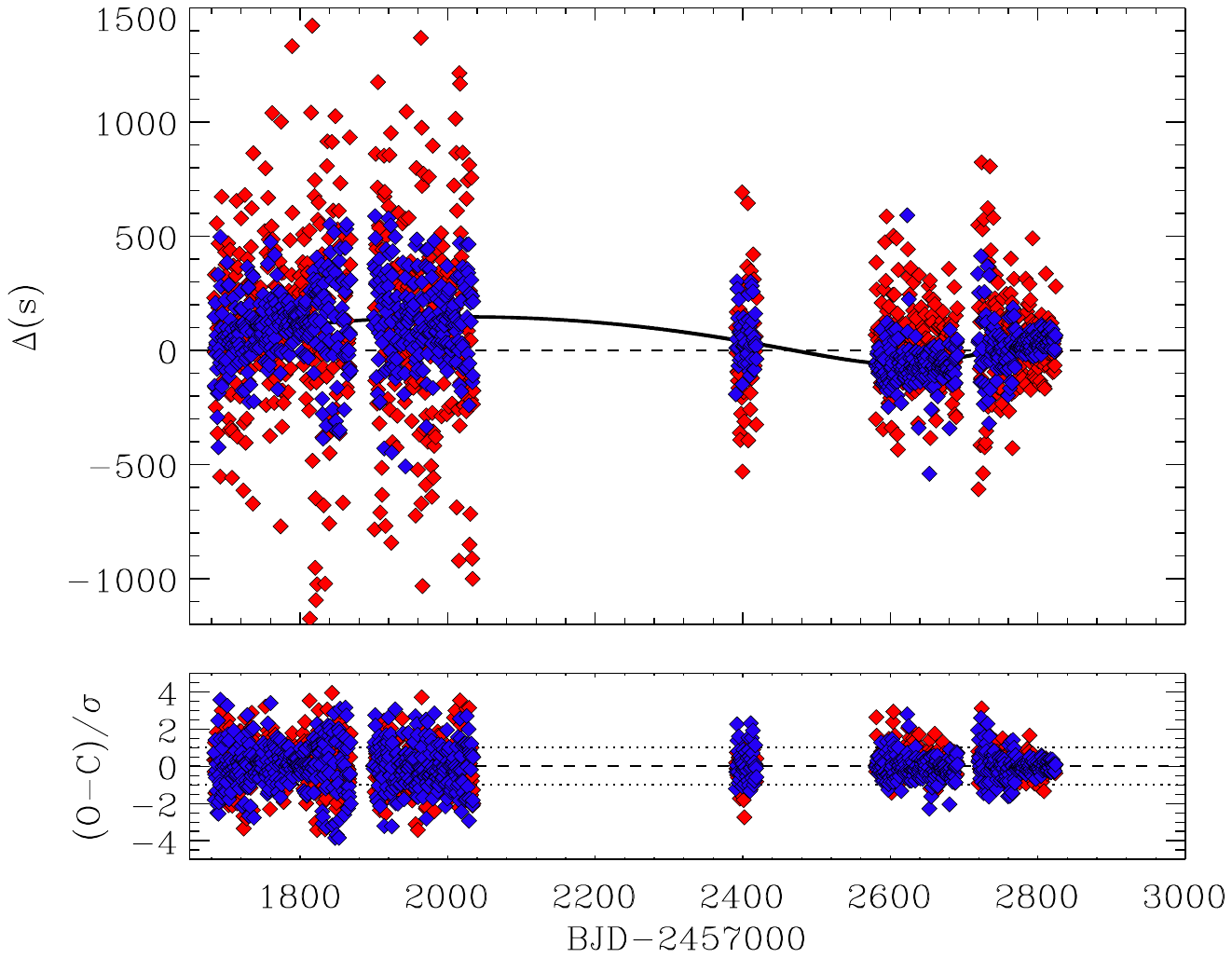}
    \caption{Same as Figure~\ref{fig:ETV_curve} but for \ticone{}.}
    \label{fig:ETV_tic21}
\end{figure}

\begin{figure}[h!]
    \centering
    \includegraphics[trim = 2.0cm 3.7cm 3.5cm 1.5cm,clip,width=0.6\columnwidth,angle=0]{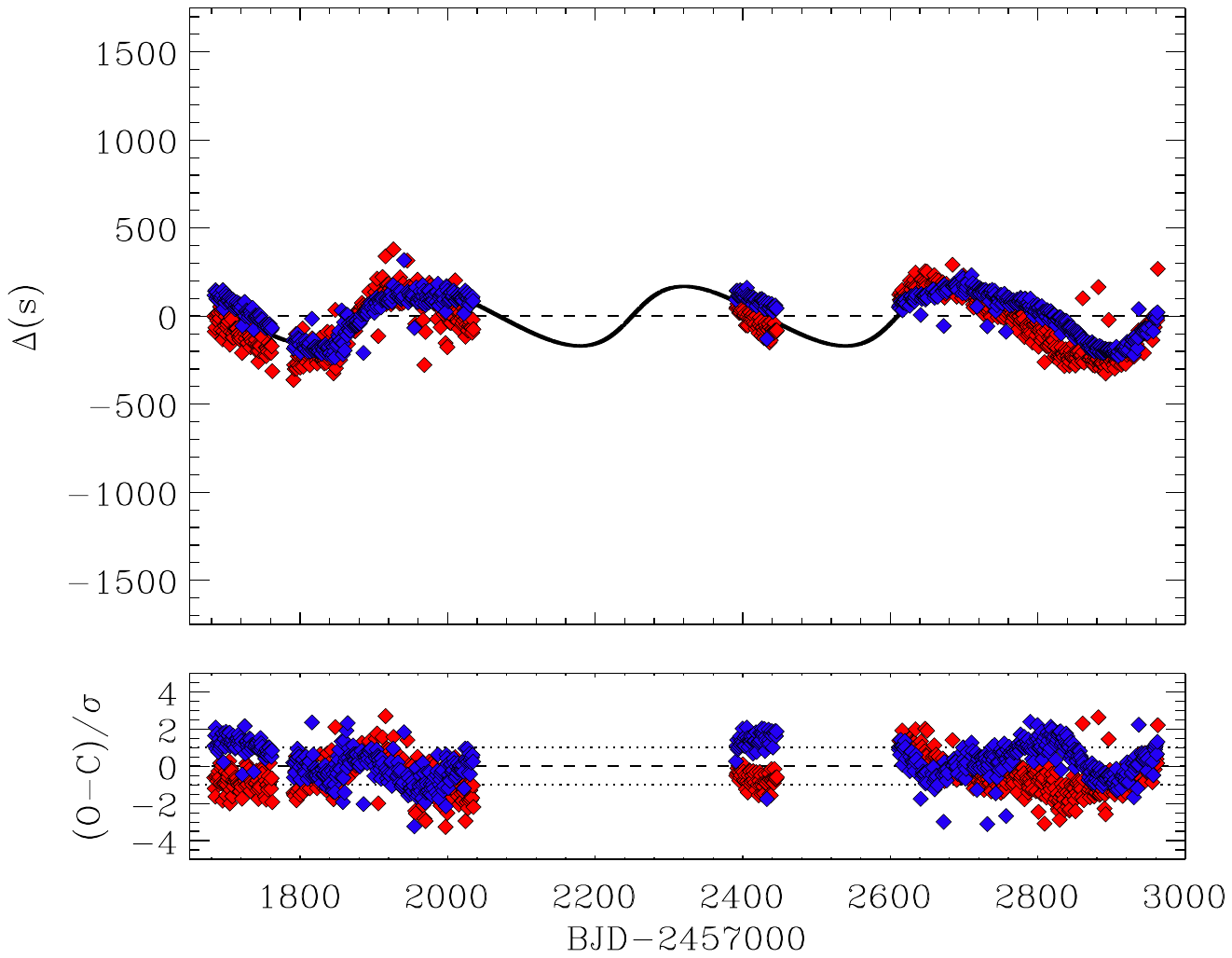}
    \includegraphics[trim = 2.0cm 2.0cm 3.5cm 1.5cm,clip,width=0.6\columnwidth,angle=0]{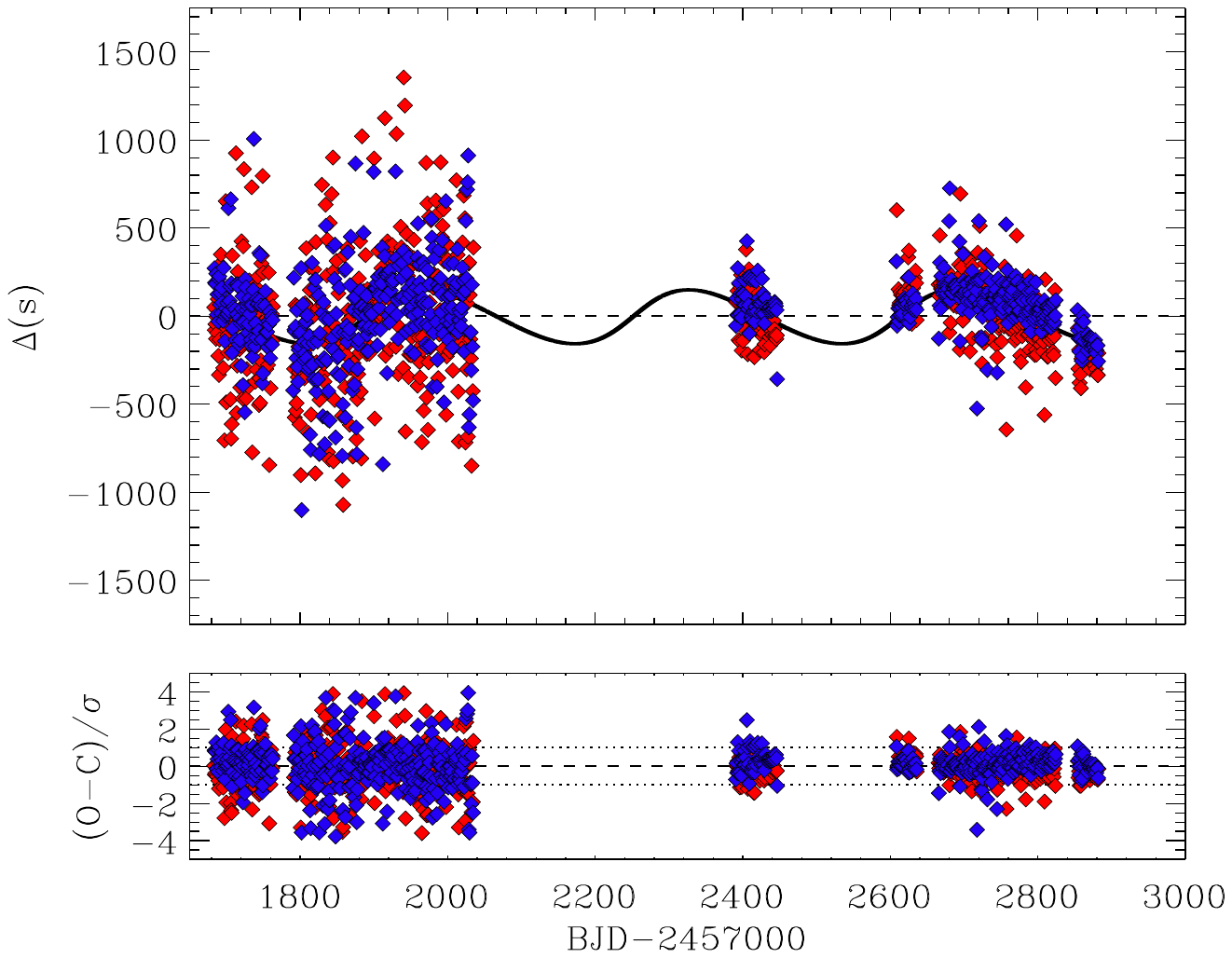}
    \caption{Same as Figure~\ref{fig:ETV_curve} but for \tictwo{}.}
    \label{fig:ETV_tic22}
\end{figure}

\begin{figure}[h!]
    \centering
    \includegraphics[trim = 2.0cm 3.7cm 3.5cm 1.5cm,clip,width=0.6\columnwidth,angle=0]{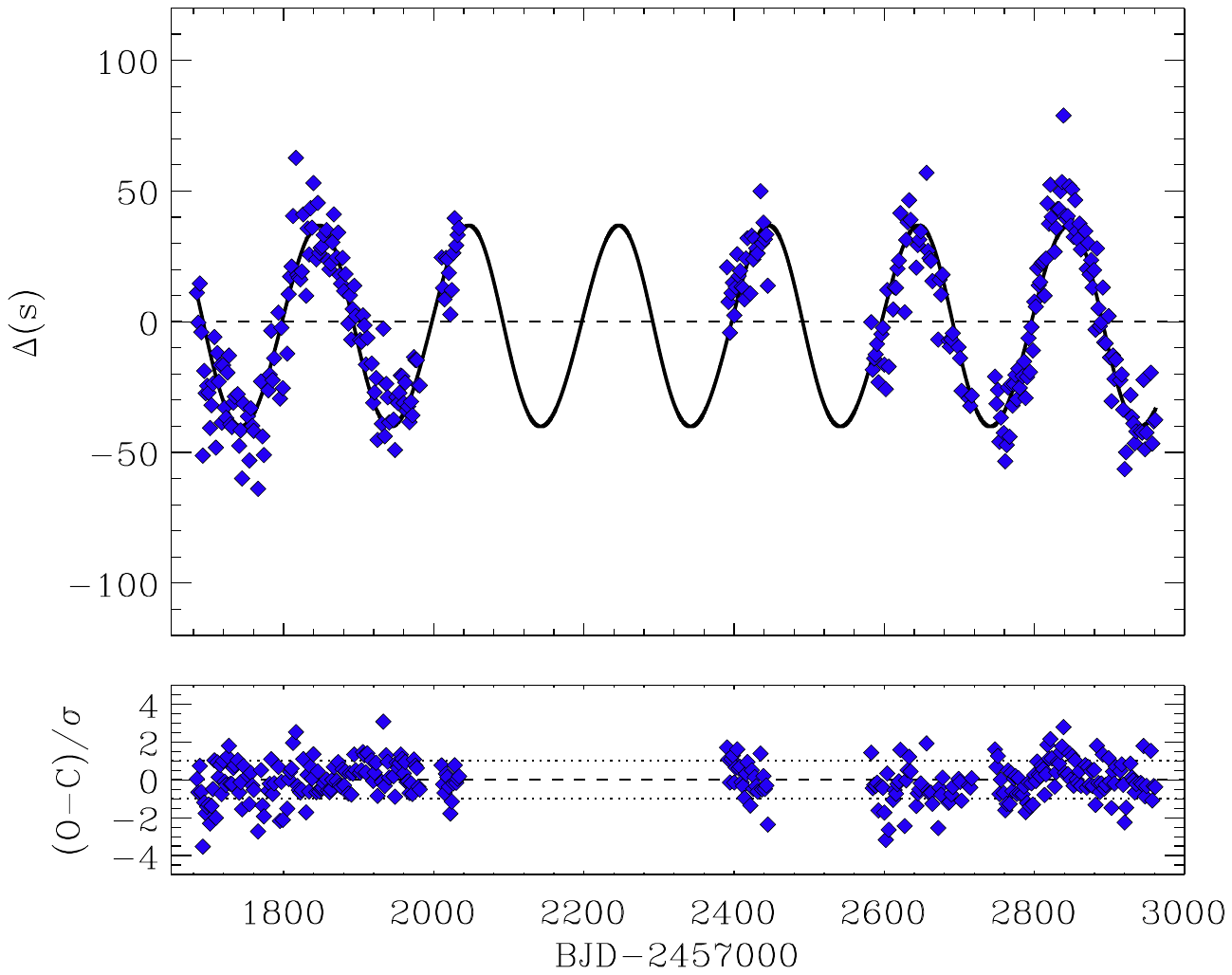}
    \includegraphics[trim = 2.0cm 2.0cm 3.5cm 1.5cm,clip,width=0.6\columnwidth,angle=0]{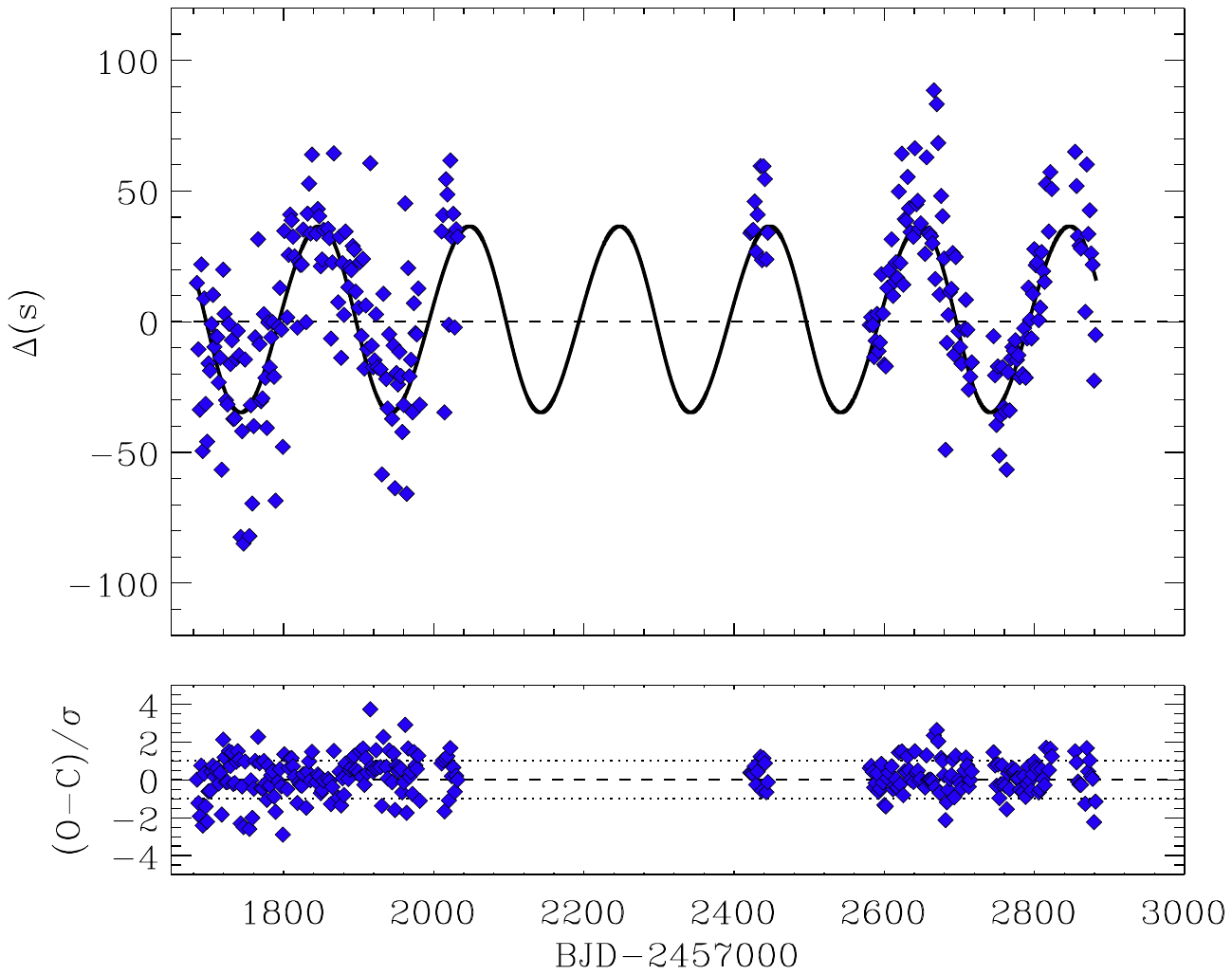}
    \caption{Same as Figure~\ref{fig:ETV_curve} but for \ticthree{}.}
\end{figure}

\begin{figure}[h!]
    \centering
    \includegraphics[trim = 2.0cm 3.7cm 3.5cm 1.5cm,clip,width=0.6\columnwidth,angle=0]{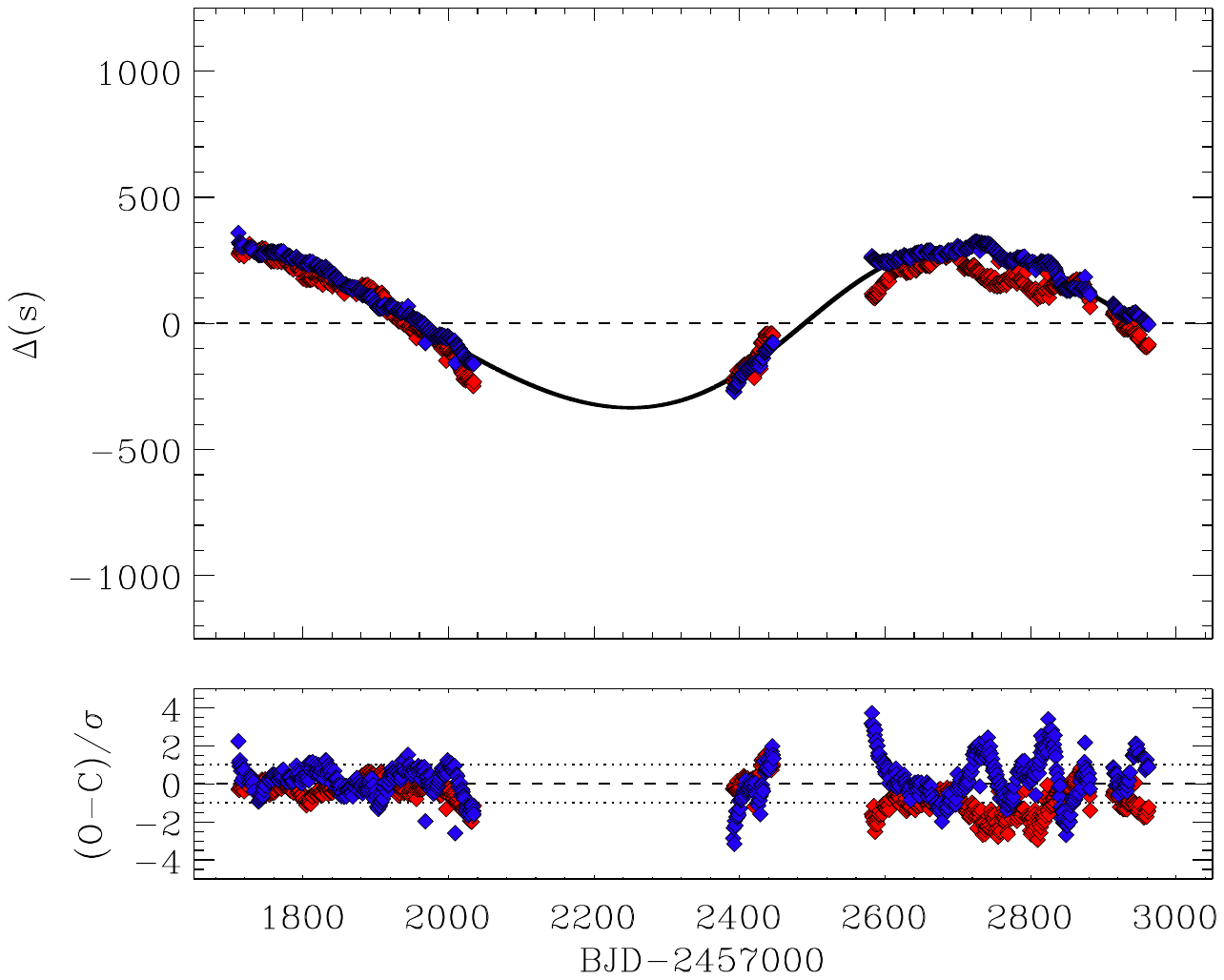}
    \includegraphics[trim = 2.0cm 2.0cm 3.5cm 1.5cm,clip,width=0.6\columnwidth,angle=0]{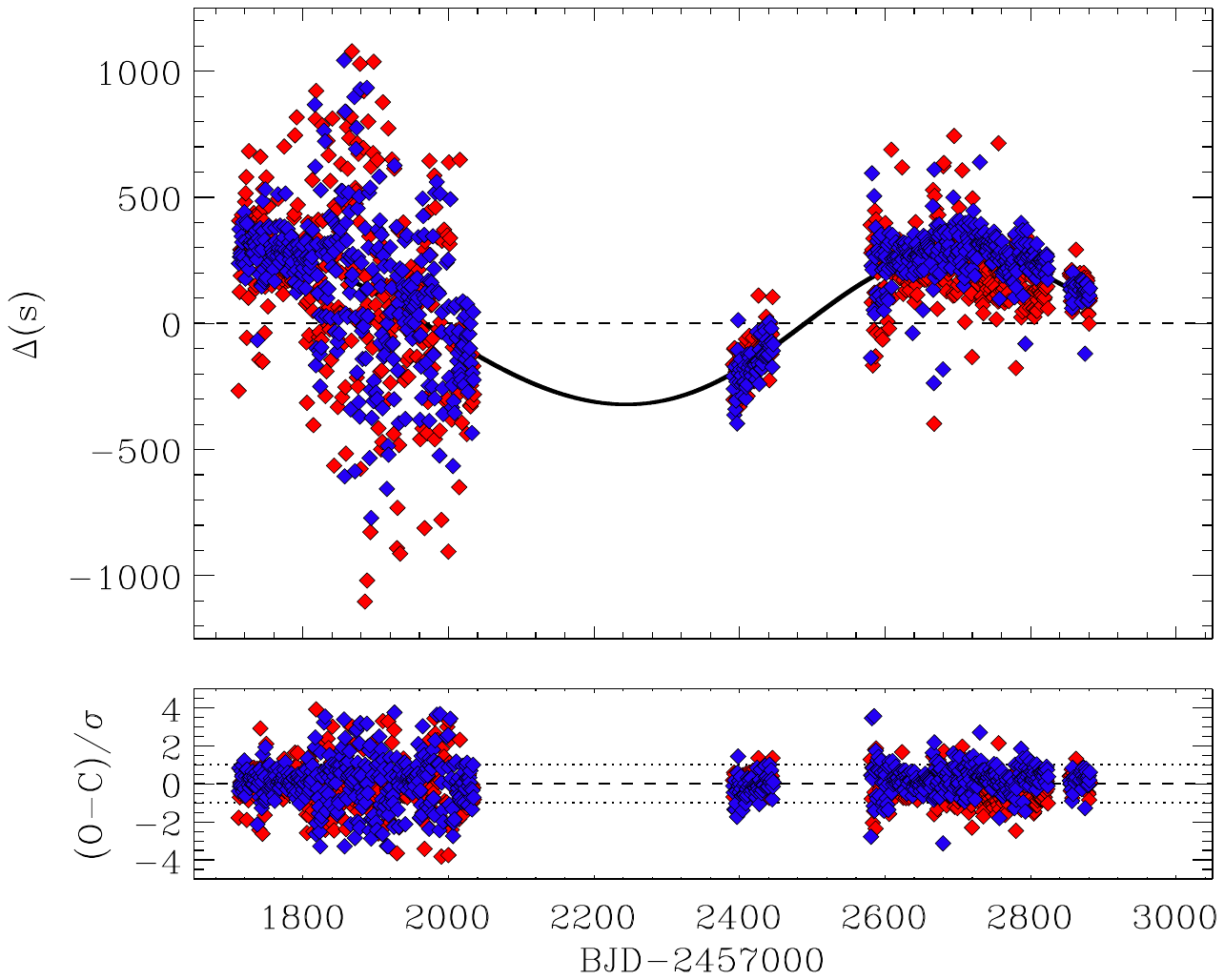}
    \caption{Same as Figure~\ref{fig:ETV_curve} but for \ticfive{}.}
    \label{fig:ETV_tic42}
\end{figure}

\clearpage
\section{Tables of orbital parameters}\label{sec:ETV_param}

\begin{table}[h!]
    \centering
    \begin{minipage}{130mm}
        \caption{Best-fit orbital parameters for \ticone{} obtained from the SC data.}
        {
        \renewcommand{\arraystretch}{1.0}
        \begin{tabular}{@{}lccc@{}}
            \hline
            \hline
                                                               &                  & 84 per cent      & 16 per cent      \\
            Parameter                                          & Median           & interval         & interval         \\
            \hline
            $P_{\rm A}$ (d)                                    & $\mbf{0.515\,2627156}$  & $\mbf{+0.000\,0000076}$ & $\mbf{-0.000\,0000071}$ \\
            $T_0$ (BJD$-245\,7000$)                            & $\mbf{1683.749\,7217}$  & $\mbf{+0.000\,0060}$    & $\mbf{-0.000\,0096}$    \\
            $A_{\rm LTTE}$ (s)                                 & \tbf{113.04}            & \tbf{+0.84}             & \tbf{$-$0.71}           \\
            $K_{\rm A}$ (km\,s$^{-1}$)                         & \tbf{2.478}             & \tbf{+0.018}            & \tbf{$-$0.021}          \\
            $P_{\rm AB}$ (d)                                   & \tbf{1159.4}            & \tbf{+3.3}              & \tbf{$-$2.1}            \\
            $T_{\rm AB}$ (BJD$-245\,7000$)                     & \tbf{2652.1}            & \tbf{+2.7}              & \tbf{$-$6.2}            \\
            $e_{\rm AB}$                                       & \tbf{0.5055}            & \tbf{+0.0029}           & \tbf{$-$0.0055}         \\
            $\omega_{\rm AB}$ ($^\circ$)                       & \tbf{283.0}             & \tbf{+1.1}              & \tbf{$-$2.5}            \\
            $a_{\rm A}\sin i_{\rm AB}$ (au)                    & \tbf{0.2280}            & \tbf{+0.0014}           & \tbf{$-$0.0014}         \\
            $f\!$($M_{\rm B}$) (M$_\odot$)                     & $\mbf{0.001\,175}$      & $\mbf{+0.000\,024}$     & $\mbf{-0.000\,021}$     \\
            $M_{\rm B} \, (i_{\rm AB}\!=90^\circ$) (M$_\odot$) & \tbf{0.1773}            & \tbf{+0.0015}           & \tbf{$-$0.0014}         \\
            \boldm{$a_{\rm AB}$} \tbf{(mas)}                   & \tbf{4.9}               & \tbf{+0.1}              & \tbf{$-$0.1}            \\
            ${\mathcal A}_{\rm dyn}/{\mathcal A}_{\rm LTTE}$   & $\mbf{0.003\,530}$      & $\mbf{+0.000\,032}$     & $\mbf{-0.000\,052}$     \\
            \hline 
        \end{tabular}
        }
    \end{minipage}
    \tablecomments{The values of $T_0$ and $P_{\rm A}$ were corrected by $c_0$ and $c_1$, respectively. The mass of the third component, $M_{\rm B}$, \tbf{the semi-major axis of the outer relative orbit,} \boldm{$a_{\rm AB}$}, and the dynamical amplitude, ${\mathcal A}_{\rm dyn}$, were computed assuming $i_{\rm AB}\!=90^\circ$ and $M_{\rm Aa} = M_{\rm Ab} = 1.00 \pm 0.01 \, {\rm M}_\odot$ (see the text).}
\end{table}

\begin{table}[h!]
    \centering
    \begin{minipage}{130mm}
        \caption{Best-fit orbital parameters for \tictwo{} obtained from the SC data.}
        \label{tab:ETV_param_tic22}
        {
        \renewcommand{\arraystretch}{1.0}
        \begin{tabular}{@{}lccc@{}}
            \hline
            \hline
                                                               &                 & 84 per cent      & 16 per cent      \\
            Parameter                                          & Median          & interval         & interval         \\
            \hline
            $P_{\rm A}$ (d)                                    & $0.820\,953946$ & $+0.000\,000043$ & $-0.000\,000049$ \\
            $T_0$ (BJD$-245\,7000$)                            & $1931.067\,665$ & $+0.000\,016$    & $-0.000\,041$    \\
            $A_{\rm LTTE}$ (s)                                 & 179.7           & +5.3             & $-$5.2           \\
            $K_{\rm A}$ (km\,s$^{-1}$)                         & 11.6            & +0.6             & $-$0.5           \\
            $P_{\rm AB}$ (d)                                   & 359.8           & +0.8             & $-$0.8           \\
            $T_{\rm AB}$ (BJD$-245\,7000$)                     & 2609.9          & +1.4             & $-$1.3           \\
            $e_{\rm AB}$                                       & 0.341           & +0.067           & $-$0.050         \\
            $\omega_{\rm AB}$ ($^\circ$)                       & 0.25            & +0.39            & $-$0.18          \\
            $a_{\rm A}\sin i_{\rm AB}$ (au)                    & 0.360           & +0.011           & $-$0.010         \\
            $f\!$($M_{\rm B}$) (M$_\odot$)                     & 0.0481          & +0.0044          & $-$0.0041        \\
            $M_{\rm B} \, (i_{\rm AB}\!=90^\circ$) (M$_\odot$) & 0.706           & +0.026           & $-$0.025         \\
            \boldm{$a_{\rm AB}$} \tbf{(mas)}                   & \tbf{19.4}      & \tbf{+0.2}       & \tbf{$-$0.2}     \\
            ${\mathcal A}_{\rm dyn}/{\mathcal A}_{\rm LTTE}$   & 0.0480          & +0.0056          & $-$0.0033        \\
            \hline 
        \end{tabular}
        }
    \end{minipage}
    \tablecomments{The values of $T_0$ and $P_{\rm A}$ were corrected by $c_0$ and $c_1$, respectively. The mass of the third component, $M_{\rm B}$, \tbf{the semi-major axis of the outer relative orbit,} \boldm{$a_{\rm AB}$}, and the dynamical amplitude, ${\mathcal A}_{\rm dyn}$, were computed assuming $i_{\rm AB}\!=90^\circ$ and $M_{\rm Aa} = M_{\rm Ab} = 1.00 \pm 0.01 \, {\rm M}_\odot$ (see the text).}
\end{table}

\begin{table}[h!]
    \centering
    \begin{minipage}{130mm}
        \caption{Best-fit orbital parameters for \ticthree{} obtained from the SC data.}
        {
        \renewcommand{\arraystretch}{1.0}
        \begin{tabular}{@{}lccc@{}}
            \hline
            \hline
                                                               &                 & 84 per cent      & 16 per cent      \\
            Parameter                                          & Median          & interval         & interval         \\
            \hline
            $P_{\rm A}$ (d)                                    & $1.939\,651524$ & $+0.000\,000046$ & $-0.000\,000044$ \\
            $T_0$ (BJD$-245\,7000$)                            & $1901.860\,271$ & $+0.000\,031$    & $-0.000\,052$    \\
            $A_{\rm LTTE}$ (s)                                 & 38.6            & +1.3             & $-$1.3           \\
            $K_{\rm A}$ (km\,s$^{-1}$)                         & 4.2             & +0.1             & $-$0.1           \\
            $P_{\rm AB}$ (d)                                   & 199.6           & +0.5             & $-$0.5           \\
            $T_{\rm AB}$ (BJD$-245\,7000$)                     & 2675.7          & +20.0            & $-$8.4           \\
            $e_{\rm AB}$                                       & 0.079           & +0.062           & $-$0.049         \\
            $\omega_{\rm AB}$ ($^\circ$)                       & 147.9           & +35.4            & $-$14.9          \\
            $a_{\rm A}\sin i_{\rm AB}$ (au)                    & 0.0773          & +0.0027          & $-$0.0027        \\
            $f\!$($M_{\rm B}$) (M$_\odot$)                     & 0.00154         & +0.00016         & $-$0.00015       \\
            $M_{\rm B} \, (i_{\rm AB}\!=90^\circ$) (M$_\odot$) & 0.1953          & +0.0072          & $-$0.0072        \\
            \boldm{$a_{\rm AB}$} \tbf{(mas)}                   & \tbf{4.560}     & \tbf{+0.019}     & \tbf{$-$0.019}   \\
            ${\mathcal A}_{\rm dyn}/{\mathcal A}_{\rm LTTE}$   & 0.605           & +0.014           & $-$0.006         \\
            \hline 
        \end{tabular}
        }
    \end{minipage}
    \tablecomments{The values of $T_0$ and $P_{\rm A}$ were corrected by $c_0$ and $c_1$, respectively. The mass of the third component, $M_{\rm B}$, \tbf{the semi-major axis of the outer relative orbit,} \boldm{$a_{\rm AB}$}, and the dynamical amplitude, ${\mathcal A}_{\rm dyn}$, were computed assuming $i_{\rm AB}\!=90^\circ$ and $M_{\rm Aa} = M_{\rm Ab} = 1.00 \pm 0.01 \, {\rm M}_\odot$ (see the text).}
\end{table}

\begin{table}[h!]
    \centering
    \begin{minipage}{130mm}
        \caption{Best-fit orbital parameters for \ticfive{} obtained from the SC data.}
        \label{tab:ETV_param_tic42}
        {
        \renewcommand{\arraystretch}{1.0}
        \begin{tabular}{@{}lccc@{}}
            \hline
            \hline
                                                               &                 & 84 per cent      & 16 per cent      \\
            Parameter                                          & Median          & interval         & interval         \\
            \hline
            $P_{\rm A}$ (d)                                    & $\mbf{0.744\,836364}$  & $\mbf{+0.000\,000038}$ & $\mbf{-0.000\,000041}$ \\
            $T_0$ (BJD$-245\,7000$)                            & $\mbf{1712.098\,436}$  & $\mbf{+0.000\,017}$    & $\mbf{-0.000\,006}$    \\
            $A_{\rm LTTE}$ (s)                                 & \tbf{316.5}            & \tbf{+1.3}             & \tbf{$-$1.5}           \\
            $K_{\rm A}$ (km\,s$^{-1}$)                         & \tbf{7.134}            & \tbf{+0.039}           & \tbf{$-$0.042}         \\
            $P_{\rm AB}$ (d)                                   & \tbf{995.6}            & \tbf{+3.4}             & \tbf{$-$2.9}           \\
            $T_{\rm AB}$ (BJD$-245\,7000$)                     & \tbf{2525.3}           & \tbf{+4.3}             & \tbf{$-$4.1}           \\
            $e_{\rm AB}$                                       & \tbf{0.1745}           & \tbf{+0.0073}          & \tbf{$-$0.0087}        \\
            $\omega_{\rm AB}$ ($^\circ$)                       & \tbf{18.8}             & \tbf{+1.6}             & \tbf{$-$1.6}           \\
            $a_{\rm A}\sin i_{\rm AB}$ (au)                    & \tbf{0.6430}           & \tbf{+0.0031}          & \tbf{$-$0.0032}        \\
            $f\!$($M_{\rm B}$) (M$_\odot$)                     & \tbf{0.03578}          & \tbf{+0.00053}         & \tbf{$-$0.00055}       \\
            $M_{\rm B} \, (i_{\rm AB}\!=90^\circ$) (M$_\odot$) & \tbf{0.6274}           & \tbf{+0.0045}          & \tbf{$-$0.0046}        \\
            \boldm{$a_{\rm AB}$} \tbf{(mas)}                   & \tbf{6.426}            & \tbf{+0.090}           & \tbf{$-$0.090}         \\            
            ${\mathcal A}_{\rm dyn}/{\mathcal A}_{\rm LTTE}$   & $\mbf{0.006\,052}$     & $\mbf{+0.000\,037}$    & $\mbf{-0.000\,034}$    \\
            \hline 
        \end{tabular}
        }
    \end{minipage}
    \tablecomments{The values of $T_0$ and $P_{\rm A}$ were corrected by $c_0$ and $c_1$, respectively. The mass of the third component, $M_{\rm B}$, \tbf{the semi-major axis of the outer relative orbit,} \boldm{$a_{\rm AB}$}, and the dynamical amplitude, ${\mathcal A}_{\rm dyn}$, were computed assuming $i_{\rm AB}\!=90^\circ$ and $M_{\rm Aa} = M_{\rm Ab} = 1.00 \pm 0.01 \, {\rm M}_\odot$ (see the text).}
\end{table}

\clearpage
\bibliography{fredm}{}
\bibliographystyle{aasjournal}



\end{document}